\documentclass[
 reprint,
 showpacs, preprintnumbers,
 amsmath, amssymb,
 aps,
 pra,
 floatfix,
 superscriptaddress
]{revtex4-1}

\usepackage{graphicx}
\usepackage{tikz}
\usepackage{xcolor}
\usetikzlibrary{arrows}

\makeatletter
\newcommand{\colorcaption}[2][]{%
  \begingroup%
  \renewcommand{\@caption@fignum@sep}{ (Color online). }%
  \caption[#1]{#2}%
  \endgroup%
}
\makeatother

\makeatletter
\def\maketag@@@#1{\hbox{\m@th\normalfont\normalsize#1}}
\makeatother

\usepackage{dcolumn}
\usepackage{bm}
\usepackage{bbm}
\usepackage{upgreek}
\usepackage{hyperref}
\usepackage{blindtext}
\usepackage{hyphsubst}

\usepackage{changes}

\usepackage{dsfont}

\usepackage{newtxtext,newtxmath}

\definecolor{darkgreen}{RGB}{0 100 0}

\begin{document}

	\preprint{}


	\title{Microscopic study of the Josephson supercurrent diode effect in Josephson junctions \newline based on two-dimensional electron gas}
	
	\author{Andreas Costa}%
	\email[Corresponding author: ]{andreas.costa@physik.uni-regensburg.de}
 	\affiliation{Institute for Theoretical Physics, University of Regensburg, 93040 Regensburg, Germany}

 	\author{Jaroslav Fabian}%
 	\affiliation{Institute for Theoretical Physics, University of Regensburg, 93040 Regensburg, Germany}

    \author{Denis Kochan}%
 	\affiliation{Institute of Physics, Slovak Academy of Sciences, 84511 Bratislava, Slovakia}
    \affiliation{Institute for Theoretical Physics, University of Regensburg, 93040 Regensburg, Germany}

 	\date{\today}

    
    \begin{abstract}
    	    Superconducting systems that simultaneously lack space-inversion and time-reversal symmetries have recently been the subject of a flurry of experimental and theoretical research activities. 
            Their ability to carry supercurrents with magnitudes depending on the polarity~(current direction)---termed the supercurrent diode effect---might be practically exploited to design dissipationless counterparts of contemporary semiconductor-based diodes. 
            Magnetic Josephson junctions realized in the two-dimensional electron gas~(2DEG) within a narrow quantum well through proximity to conventional superconductors perhaps belong to the most striking and versatile platforms for such supercurrent rectifiers. 
            Starting from the Bogoliubov--de Gennes approach, we provide a minimal theoretical model to explore the impact of the spin-orbit coupling and magnetic exchange inside the 2DEG on the Andreev bound states and Josephson current--phase relations. 
            Assuming realistic junction parameters, we evaluate the polarity-dependent critical currents to quantify the efficiency of these Josephson junctions as supercurrent diodes, and discuss the tunability of the Josephson supercurrent diode effect in terms of spin-orbit coupling, magnetic exchange, and transparency of the nonsuperconducting weak link. 
            Furthermore, we demonstrate that the junctions might undergo current-reversing $ 0 $--$ \pi $-like phase transitions at large enough magnetic exchange, which appear as sharp peaks followed by a sudden suppression in the supercurrent-diode-effect efficiency. 
            The characteristics of the Josephson supercurrent diode effect obtained from our model convincingly reproduce many unique features observed in recent experiments, validating its robustness and suitability for further studies.           
    \end{abstract}
    
    
    \maketitle
    

    \section{Introduction   \label{Sec_Introduction}}

    Since their initial experimental discovery in synthesized (Nb/Ta/V)$ _n $-based superlattices~\cite{Ando2020}, nonreciprocal supercurrent phenomena have attracted enormous interest and have been thoroughly investigated in various systems covering, e.g., twisted magic-angle bi- and trilayer graphene~\cite{DiezMerida2023,lin2022zerofield,scammell2022theory}, proximitized topological insulators~\cite{BoLu2022,Pal2022,Fu2022TopoDiode} and van~der~Waals heterostructures~\cite{Wu2022,Bauriedl2022}, as well as Josephson junctions~\cite{Baumgartner2021,Baumgartner2022a,Baumgartner2022,Jeon2022,Pal2022,Turini2022,FrolovQW2022,Costa2023,*Costa2023NatNano,Banerjee2023}. 
    Their ability to carry supercurrents of different magnitude depending on the polarity, which motivated terming them \emph{supercurrent diodes}~\cite{Hu2007}, makes these systems promising candidates for future applications in dissipation-free electronics and information technology. 

    Nonreciprocal supercurrents are most commonly induced combining the spin-orbit~coupling~(SOC)~\cite{Fabian2004,Fabian2007} of a space-inversion-symmetry-lacking material with a simultaneously time-reversal-symmetry-breaking magnetic exchange interaction~\cite{Ando2020,Baumgartner2022a,Baumgartner2022,Wu2022,Jeon2022,Pal2022,Bauriedl2022,Turini2022,FrolovQW2022,Kochan2023}, such as the Zeeman coupling triggered by an appropriately aligned external magnetic field or an intrinsic exchange splitting. 
    Detecting the supercurrent~diode~effect~(SDE) might therefore serve as a probe of broken space-inversion and time-reversal symmetries, which is often accumulated in the common name noncentrosymmetry~\cite{Edelstein1989,*Edelstein1989alt,Edelstein1996,Dimitrova2007,Mineev2012}.  
        
    From the theoretical point of view, different microscopic and phenomenological explanations of the SDE have been proposed. 
    The first approach~\cite{Daido2022,Yuan2022,He2022,Ilic2022,Davydova2022,Banerjee2023}, elaborating on the understanding of the SDE in quasi-2D superconductors, relied on a shift of the SOC-split Fermi surfaces through the Zeeman coupling, which resulted in a finite Cooper-pair momentum and thus in a helical phase that revealed a nonreciprocal current response. The second one~\cite{Buzdin2008,Grein2009,Baumgartner2022a,Baumgartner2022,Costa2023,*Costa2023NatNano,Lotfizadeh2023} linked the SDE in Josephson junctions to a strong phase-asymmetry of the Andreev-bound-state spectra~\cite{Andreev1966,*Andreev1966alt}, which likewise originated from the competition of SOC and Zeeman coupling. 
    Since the coherent Cooper-pair transport through Josephson junctions is mediated by electrons tunneling via these Andreev bound states, the phase-asymmetry will be imprinted on the Josephson current--phase~relation~(CPR) and endow the CPR with an anomalous $ \varphi_0 $-phase shift~\cite{Bezuglyi2002,Krive2004,Buzdin2008,Reynoso2008,Zazunov2009,Liu2010a,Liu2010,Liu2011,Reynoso2012,YokoyamaJPSJ2013,Brunetti2013,Shen2014,Yokoyama2014Anomalous,Konschelle2015,Szombati2016,Assouline2019,Mayer2020b,Strambini2020}, the SDE~\cite{Grein2009,Lotfizadeh2023}, and even current-reversing $0$--$\pi$-like phase transitions~\cite{Bulaevskii1977a,*Bulaevskii1977b,Ryazanov2001,Andersen2006,Kawabata2010,Kawabata2012,YokoyamaJPSJ2013,Yokoyama2014Anomalous}. 
    Moreover, a phenomenological account of the SDE in both thin films and Josephson junctions was provided in~Ref.~\cite{Kochan2023} by means of a generalized Ginzburg--Landau functional tailored to noncentrosymmetric interactions triggered within the superconducting condensate.
    
    Analytical calculations~\cite{Buzdin2008} suggested that $ |\varphi_0| $ scales, in the simplest case, linearly with the product of the SOC and Zeeman-coupling strengths, and inversely with the square of the velocity components of the incident electrons~(holes) that are parallel to the current direction. 
    Since these velocity components are different for different transverse channels of two-dimensional junctions, the CPRs of the individual channels acquire distinct $ \varphi_0 $-shifts. 
    In multi-channel Josephson junctions, the total CPR emerges as the sum of all these channels' CPRs and therefore gets strongly distorted. 
    Its asymmetry and anharmonicity with respect to the superconducting phase difference~$\varphi$ finally results in different magnitudes of the positive and negative critical currents as a clear fingerprint of the (DC) Josephson SDE, on which we will focus in this work. 

    As a brief comment, apart from the $\varphi_0$-shifts and polarity-dependent critical currents that can be probed by DC measurements, also the inflection point of the total CPR will be shifted, which can be probed in an AC regime and leads to the magnetochiral anisotropy of the Josephson inductance, usually referred to as AC SDE~\cite{Baumgartner2021,Baumgartner2022a,Baumgartner2022,Costa2023,*Costa2023NatNano}. 
    However, both the AC and DC SDEs can be undoubtedly explained in terms of a single theory that gives access to the full CPR of the junction. 

    In this paper, we present such a detailed theoretical study that illustrates the ramifications of the Rashba SOC~\cite{Bychkov1984,Bychkov1984b,*Bychkov1984c} and magnetic exchange on the Andreev-bound-state spectra, and hence on the therefrom resulting Josephson-transport characteristics. We focus on ballistic two-dimensional multi-channel Josephson junctions with a simple, but interaction-wise complex, model Hamiltonian that promotes a
    sizable and robust (DC) Josephson SDE, along with pronounced global $\varphi_0$-shifts in the total CPRs and current-reversing $ 0 $--$ \pi $-like~transitions. 
    We demonstrate that these phenomena are connected such that $ 0 $--$ \pi $-like transitions can be identified by means of the critical-current asymmetries that are imprinted on the SDE efficiencies. Moreover, we explore the tunability of the SDE in terms of SOC, magnetic exchange, and transparency of the junction, assuming experimentally realistic parameters, and reproduce all characteristic SDE features observed in recent experiments~\cite{Baumgartner2022a,Baumgartner2022,Costa2023,*Costa2023NatNano} on Al-gated InAs 2DEG-based Josephson junctions in an external in-plane magnetic field.

    The paper is organized in the following way. 
    In Sec.~\ref{Sec_Model}, we elaborate on the theoretical model that we apply to compute the Andreev bound states and CPRs of the 2DEG-based Josephson junctions. The CPRs are crucial to extract the polarity dependence of the critical current and to quantify the SDE. 
    We discuss and thoroughly analyze the results of our numerical calculations in~Sec.~\ref{Sec_Results}, with a special emphasis on the modifications the Andreev spectra and CPRs experience in the presence of magnetic exchange~(Secs.~\ref{Sec_BoundStates_Nonmagnetic} and \ref{Sec_BoundStates_Magnetic}), and on the Josephson SDE~(Sec.~\ref{Sec_Diode}). 
    We briefly conclude our analyses and comment on our results in~Sec.~\ref{Sec_Conclusions}.

    \section{Theoretical model  \label{Sec_Model}} 
    
\begin{figure}
        \centering
        \includegraphics[width=0.49\textwidth]{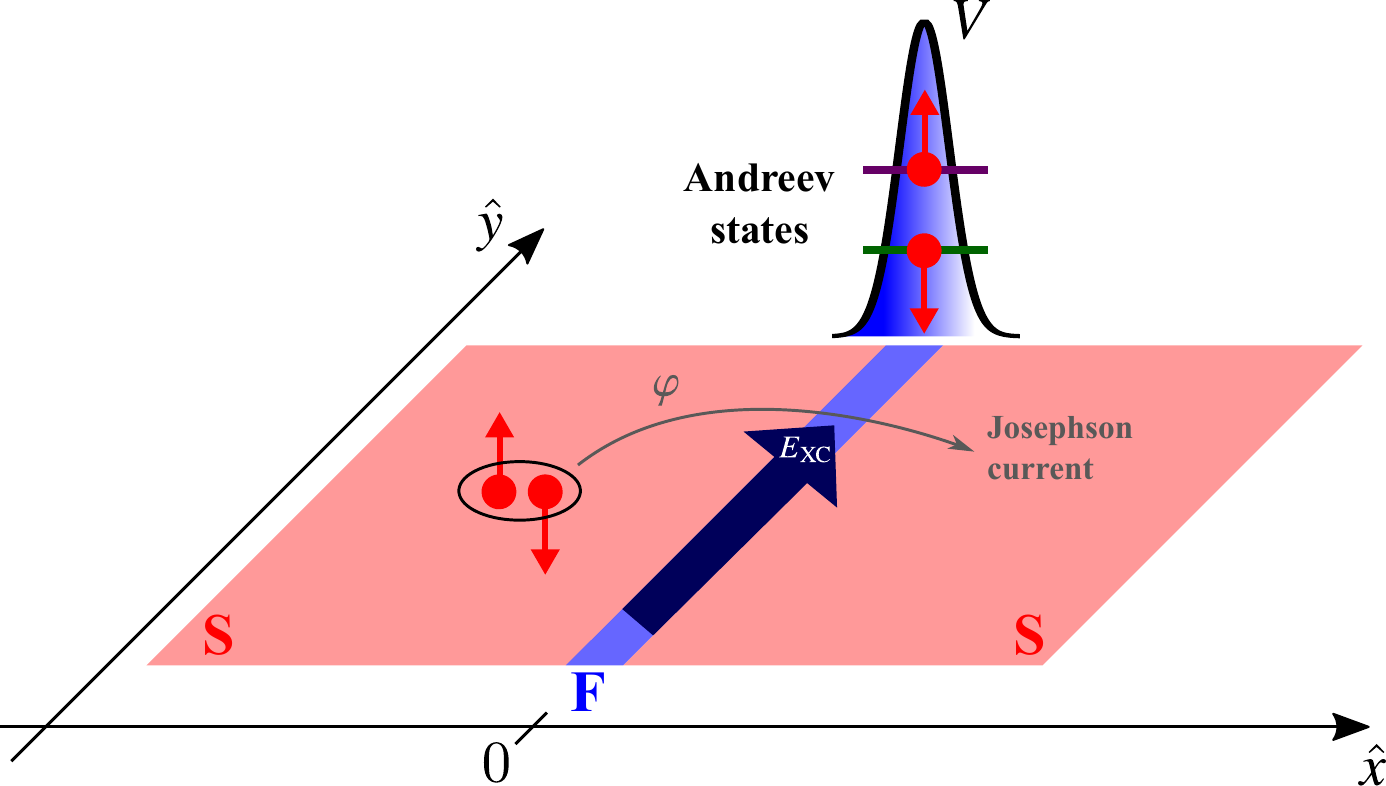}
        \caption{2DEG-based Josephson junction, with the transverse width~$W$, consisting of two semi-infinite proximitized superconducting~(S) regions ($x<0$ and $x>0$) and the
        delta-like ferromagnetic~(F) tunnel barrier at $x=0$. The latter introduces the tunnel potential~$ V $ to modify the transparency of the junction link and additionally accounts for the in-plane magnetic exchange field $E_{\mathrm{XC}}$ that acts on the spin degrees and is aligned perpendicular to the direction of the Josephson current. The current is triggered by a superconducting phase difference~$ \varphi $ along the junction, which induces the tunneling of Cooper pairs via the discrete Andreev bound states that are localized in the F~region. The width~$ W $ of the junction controls the number of the transverse Andreev modes~(channels).} 
        \label{Fig_System}
    \end{figure}

    We consider a ballistic superconductor/ferro\-mag\-net/su\-per\-con\-ductor~(S/F/S) Josephson~junction realized in a two-dimensional electron~gas~(2DEG) that is proximitized by an $ s $-wave superconductor and driven by Rashba SOC. As an example, one could consider a narrow InAs quantum well with epitaxially grown aluminium top gates~\cite{Baumgartner2021}. The 2DEG proximitized with superconducting correlations spans the regions~$ x < 0 $ and $ x > 0 $, which are connected by a delta-like ferromagnetic tunnel barrier at $x=0$ that serves as weak link; see Fig.~\ref{Fig_System} for a schematic illustration. The barrier accounts not only for the in experiments often reduced interfacial transparencies of the junction~\cite{Baumgartner2021}, but additionally for exchange correlations due to, e.g., an intrinsic magnetic exchange splitting or Zeeman coupling triggered by an applied external magnetic field. Since we focus on the ultrathin barrier and on systems for which the Rashba SOC dominates, we neglect all magnetism-induced pair-breaking effects outside the ferromagnetic barrier.

    \subsection{Andreev bound states}

    We model the 2DEG-based S/F/S junction by means of the stationary Bogoliubov--de~Gennes~Hamiltonian~\cite{DeGennes1989} 
    \begin{equation}
        \hat{\mathcal{H}}_\mathrm{BdG} = \left[ \begin{matrix} \hat{\mathcal{H}}_\mathrm{e} & \hat{\Delta}_\mathrm{S} (x) \\ \hat{\Delta}_\mathrm{S}^\dagger (x) & \hat{\mathcal{H}}_\mathrm{h} \end{matrix} \right] , 
        \label{Eq_BdG_1}
    \end{equation}
    where 
    \begin{align}
        \hat{\mathcal{H}}_\mathrm{e} &= \left[ - \frac{\hbar^2}{2m} \left( \frac{\partial^2}{\partial x^2} + \frac{\partial^2}{\partial y^2} \right) - \mu \right] \hat{\sigma}_0 \nonumber \nonumber \\
        &\hspace{70 pt} - \mathrm{i} \alpha \left( \hat{\sigma}_x \frac{\partial}{\partial y} - \hat{\sigma}_y \frac{\partial}{\partial x} \right)  \nonumber \\
        &\hspace{70 pt} + \left( Vd \, \hat{\sigma}_0 + E_\mathrm{XC}d \,\hat{\sigma}_y \right) \, \delta(x) 
        \label{Eq_BdG_2}
    \end{align} 
    represents the single-particle Hamiltonian of electrons and $ \hat{\mathcal{H}}_\mathrm{h} = -\hat{\sigma}_y \hat{\mathcal{H}}_\mathrm{e}^* \hat{\sigma}_y $ of their hole counterparts.  Throughout the text, $ \hat{\sigma}_0 $ and $ \hat{\sigma}_i $ stand, correspondingly, for the 
    $ 2 \times 2 $ identity and the $ i $th Pauli matrices in spin space. 
    To simplify the analytical calculations, we assume that the effective masses~$ m $, as well as the Fermi~level~$ \mu $, are the same throughout the junction. Similarly, we assume a uniform strength~$ \alpha $ for the Rashba SOC that originates from the broken space-inversion symmetry. Moreover, the delta-like barrier with an effective scattering length $ d $ is characterized by the scalar potential~$ V $ and the magnetic exchange~$ E_\mathrm{XC} $ acting perpendicular to the current direction. 
    While the first accounts for the reduced junction transparency~\cite{DeJong1995,Zutic1999,Zutic2000,*Zutic2000alt,Costa2017,Costa2018,Costa2019,Costa2020,Costa2021}, the second 
    gives the two spin projections along the junction direction different energies. 
    The delta-like model enormously simplifies the modeling of the system at the analytical level~(we briefly comment on its validity in~Sec.~\ref{Sec_BoundStates_Magnetic}). However, it also needs to be mentioned that the finite thickness of the link in real junctions typically imprints a highly nontrivial angular dependence on its tunneling transparency. Earlier works demonstrated, for example, that thin~(nearly delta-like) links mostly favor the tunneling of (quasi)particles that are incident under sufficiently large angles with respect to the normal plane~(known as the ``acceptance cone'')~\cite{Fogelstrom1997,Bell1988,Bauer1993}, which means that channels with larger~$ |k_y| $ shall contribute with a larger weight to the total Josephson current, which would be itself beneficial for our model. However, for simplicity, we do not account for this effect in the present study and leave it open for its follow-up.

    Inside the superconducting regions, we assume an $ s $-wave coherence governed by the pairing potential 
    \begin{equation}
        \hat{\Delta}_\mathrm{S} (x) = \Delta_0 \left[ \Theta(-x) + \mathrm{e}^{\mathrm{i} \varphi} \Theta(x) \right] \hat{\sigma}_0 ,
    \end{equation}
    where $ \Delta_0 >0$ refers to the magnitude of the proximity-induced superconducting gap in the 2DEG and 
    $ \varphi $ to the corresponding superconducting phase difference.
    Approximating $ \hat{\Delta}_\mathrm{S} $ through a step function, without a self-consistent treatment, is the standard approximation~\cite{Halterman2001} assuming that spatial variations of the gap are developing on a length scale smaller than the coherence length.

    As the wave~vector~$ k_y $ parallel to the junction interface stays conserved, a general bound-state solution of the Bogoliubov--de~Gennes equation, $ \hat{\mathcal{H}}_\mathrm{BdG} \Psi(x,y) = E \Psi(x,y) $, with $|E|<\Delta_0$
    factorizes into 
    \begin{equation}
        \Psi(x,y) = \psi(x) \, \mathrm{e}^{\mathrm{i} k_y y} , 
        \label{Eq_WaveFunction}
    \end{equation}
    where the component along the junction, $\psi(x)$, reads 
    \begin{align}
        \psi(x < 0) &= \mathcal{A} \mathrm{e}^{-\mathrm{i} q_{x,\mathrm{e}}^\Uparrow x} \left[ \begin{matrix} \mathrm{i} \alpha'_\Uparrow u \\ u \\ \mathrm{i} \alpha'_\Uparrow v \\ v \end{matrix} \right] 
        + \mathcal{B} \mathrm{e}^{\mathrm{i} q_{x,\mathrm{h}}^\Uparrow x} \left[ \begin{matrix} \mathrm{i} \tilde{\alpha}'_\Uparrow v \\ v \\ \mathrm{i} \tilde{\alpha}'_\Uparrow u \\ u \end{matrix} \right] \nonumber \\
        &+ \mathcal{C} \mathrm{e}^{-\mathrm{i} q_{x,\mathrm{e}}^\Downarrow x} \left[ \begin{matrix} -\mathrm{i} \alpha'_\Downarrow u \\ u \\ - \mathrm{i} \alpha'_\Downarrow v \\ v \end{matrix} \right] 
        + \mathcal{D} \mathrm{e}^{\mathrm{i} q_{x,\mathrm{h}}^\Downarrow x} \left[ \begin{matrix} -\mathrm{i} \tilde{\alpha}'_\Downarrow v \\ v \\ - \mathrm{i} \tilde{\alpha}'_\Downarrow u \\ u \end{matrix} \right]
        \label{Eq_WF_Left}
        \intertext{and correspondingly} 
        \psi(x>0) &= \mathcal{E} \mathrm{e}^{\mathrm{i} q_{x,\mathrm{e}}^\Uparrow x} \left[ \begin{matrix} \mathrm{i} \alpha_\Uparrow u \mathrm{e}^{\mathrm{i} \varphi} \\ u \mathrm{e}^{\mathrm{i} \varphi} \\ \mathrm{i} \alpha_\Uparrow v \\ v \end{matrix} \right] 
        + \mathcal{F} \mathrm{e}^{-\mathrm{i} q_{x,\mathrm{h}}^\Uparrow x} \left[ \begin{matrix} \mathrm{i} \tilde{\alpha}_\Uparrow v \mathrm{e}^{\mathrm{i} \varphi} \\ v \mathrm{e}^{\mathrm{i} \varphi} \\ \mathrm{i} \tilde{\alpha}_\Uparrow u \\ u \end{matrix} \right] \nonumber \\
        & + \mathcal{G} \mathrm{e}^{\mathrm{i} q_{x,\mathrm{e}}^\Downarrow x} \left[ \begin{matrix} -\mathrm{i} \alpha_\Downarrow u \mathrm{e}^{\mathrm{i} \varphi} \\ u \mathrm{e}^{\mathrm{i} \varphi} \\ - \mathrm{i} \alpha_\Downarrow v \\ v \end{matrix} \right] 
        + \mathcal{H} \mathrm{e}^{-\mathrm{i} q_{x,\mathrm{h}}^\Downarrow x} \left[ \begin{matrix} -\mathrm{i} \tilde{\alpha}_\Downarrow v \mathrm{e}^{\mathrm{i} \varphi} \\ v \mathrm{e}^{\mathrm{i} \varphi} \\ - \mathrm{i} \tilde{\alpha}_\Downarrow u \\ u \end{matrix} \right]\,.
        \label{Eq_WF_Right} 
    \end{align}
    Contrary to the solution for the scattering states~(if $|E|>\Delta_0$), we do not need to account for incoming waves in the bound-state ansatz for $\psi(x)$. The effect of Rashba SOC, which mixes both spin projections in a $k_y$-momentum-dependent way, is included into the spinor amplitudes through the distinguished $\alpha$-labels 
    \begin{align*}
        \alpha_\Uparrow &= \frac{q_{x,\mathrm{e}}^\Uparrow - \mathrm{i} k_y}{\sqrt{\left( q_{x,\mathrm{e}}^\Uparrow \right)^2 + k_y^2}} , \quad \quad &\tilde{\alpha}_\Uparrow &= -\frac{q_{x,\mathrm{h}}^\Uparrow + \mathrm{i} k_y}{\sqrt{\left( q_{x,\mathrm{h}}^\Uparrow \right)^2 + k_y^2}} , \\
        \alpha_\Downarrow &= \frac{q_{x,\mathrm{e}}^\Downarrow - \mathrm{i} k_y}{\sqrt{\left( q_{x,\mathrm{e}}^\Downarrow \right)^2 + k_y^2}} , \quad \quad &\tilde{\alpha}_\Downarrow &= -\frac{q_{x,\mathrm{h}}^\Downarrow + \mathrm{i} k_y}{\sqrt{\left( q_{x,\mathrm{h}}^\Downarrow \right)^2 + k_y^2}} , \\
        \alpha'_\Uparrow &= -\frac{q_{x,\mathrm{e}}^\Uparrow + \mathrm{i} k_y}{\sqrt{\left( q_{x,\mathrm{e}}^\Uparrow \right)^2 + k_y^2}} , \quad \quad &\tilde{\alpha}'_\Uparrow &= \frac{q_{x,\mathrm{h}}^\Uparrow - \mathrm{i} k_y}{\sqrt{\left( q_{x,\mathrm{h}}^\Uparrow \right)^2 + k_y^2}} , \\
        \alpha'_\Downarrow &= -\frac{q_{x,\mathrm{e}}^\Downarrow + \mathrm{i} k_y}{\sqrt{\left( q_{x,\mathrm{e}}^\Downarrow \right)^2 + k_y^2}} , \quad \quad &\tilde{\alpha}'_\Downarrow &= \frac{q_{x,\mathrm{h}}^\Downarrow - \mathrm{i} k_y}{\sqrt{\left( q_{x,\mathrm{h}}^\Downarrow \right)^2 + k_y^2}}\,.
    \end{align*}
    The electron-hole content in $\psi(x)$ is encoded in the energy-dependent (and generally complex-valued) Bardeen---Cooper---Schrieffer~(BCS) coherence factors 
    \begin{align} 
        u &= \sqrt{\frac{1}{2} \left( 1 + \sqrt{1 - \frac{\Delta_0^2}{E^2}} \right)}\,, &
        v &= \sqrt{1 - u^2} \,,
        \label{Eq_BCSCoh_Fact}
    \end{align}
    while the electron-like and hole-like quasiparticle wave vectors in \emph{Andreev approximation}---i.e.,~for $ \mu \gg E , \Delta_0 $---are given by 
    \begin{align}
        q_{x,\mathrm{e}}^\Uparrow &\approx q_{x,\mathrm{h}}^\Uparrow \approx \sqrt{ k_\mathrm{F}^2 \left( \sqrt{1 + \lambda_\mathrm{SOC}^2} - \lambda_\mathrm{SOC} \right)^2 - k_y^2 } 
        \label{Eq_WV_SpinUp} 
        \intertext{and } 
        q_{x,\mathrm{e}}^\Downarrow &\approx q_{x,\mathrm{h}}^\Downarrow \approx \sqrt{ k_\mathrm{F}^2 \left( \sqrt{1 + \lambda_\mathrm{SOC}^2} + \lambda_\mathrm{SOC} \right)^2 - k_y^2 } \,.
        \label{Eq_WV_SpinDown}
    \end{align}
    For a more compact notation, we introduce the Fermi wave vector~$ k_\mathrm{F} = \sqrt{2m\mu}/\hbar $ and the dimensionless effective SOC strength~$ \lambda_\mathrm{SOC} = (m \alpha) / (\hbar^2 k_\mathrm{F}) $.

    Requiring interfacial continuity of $\psi(x)$, 
    \begin{equation}
        \psi(x) \Bigr|_{x=0_-} = \psi(x) \Bigr|_{x=0_+} ,
        \label{Eq_Boundary_Left}
    \end{equation}
    and the delta-function-characteristic jump in the first derivative  
    \begin{align}
        \frac{\hbar^2}{2md} \left(
        \frac{\partial}{\partial x} \psi(x) \Bigr|_{x=0_+}\right. &- \left.\frac{\partial}{\partial x} \psi(x) \Bigr|_{x=0_-}
        \right) 
        \label{Eq_Boundary_Right} \\
        &= \left[V - E_\mathrm{XC}
        \left(\begin{matrix} 
        0 & -\mathrm{i} & 0 & 0  \\ 
        \mathrm{i} & 0 & 0 & 0   \\ 
        0 & 0 & 0 & \mathrm{i} \\ 
        0 & 0 & -\mathrm{i} & 0
        \end{matrix}\right)
        \right] \psi(x) \Bigr|_{x=0_-} \nonumber
    \end{align}
    yields a homogeneous system of algebraic equations for the unknown wave-function amplitudes~$ \mathcal{A} , \ldots, \mathcal{H} $. 
    Requiring its nontrivial solution leads to the secular equation for the Andreev-bound-state energies~$ E $ that we solve numerically. 
    After determining $ E $, the above algebraic system---along with the normalization condition for $\psi(x)$---fully specifies the amplitudes~$ \mathcal{A} , \ldots , \mathcal{H} $ in terms of a single amplitude, say, $ \mathcal{E} $, which we choose to be real. 
    As convenient for the Blonder---Tinkham---Klapwijk~formalism~\cite{Blonder1982}, we define the dimensionless 
    $ Z = (2mVd)/(\hbar^2 k_\mathrm{F}) $ and $ \lambda_\mathrm{XC} = (2mE_\mathrm{XC}d)/(\hbar^2 k_\mathrm{F}) $ parameters that 
    quantify, respectively, the strengths of the corresponding interactions~(scalar and exchange) with respect to the Fermi energy. 
    For example, tunneling through metal/superconductor~interfaces possessing $ Z = 0.5 $ corresponds to the interfacial transparency 
    of~$ \overline{\tau} = 1 / [1 + (Z/2)^2] \approx 0.94 $, which agrees well with the recent experimental results in InAs 2DEG-based Josephson~junctions~\cite{Baumgartner2021}. If not otherwise stated, we therefore use $Z=0.5$ for all calculations below. 
    
    Continuity of $\psi(x)$ at $x=0$ allows us to express the amplitudes entering $\psi(x<0)$ 
    in terms of the amplitudes of $\psi(x>0)$,
    \begin{equation}
        (\mathcal{A},\mathcal{B},\mathcal{C},\mathcal{D})=(\mathcal{E},\mathcal{F},\mathcal{G},\mathcal{H})\cdot\mathbf{M} , 
        \label{Eq_Coeff_Left_1}
    \end{equation}
    where the entries of $\mathbf{M}$, depending on $k_y$ and $\varphi$, follow after some algebra 
    from Eqs.~\eqref{Eq_WF_Left},~\eqref{Eq_WF_Right}, and~\eqref{Eq_Boundary_Left}. 
    Analogously, the boundary condition for the derivatives of $\psi(x)$, Eq.~\eqref{Eq_Boundary_Right}, leads to another 
    set of equations, which---after substituting Eq.~\eqref{Eq_Coeff_Left_1}---can be rewritten as 
    \begin{equation}
        0=(\mathcal{E},\mathcal{F},\mathcal{G},\mathcal{H})\cdot\mathbf{N} , 
        \label{Eq_Coeff_Left_2}
    \end{equation}
    where the entries of $\mathbf{N}$ follow again from standard, but cumbersome, algebra. 
    Reorganizing Eq.~\eqref{Eq_Coeff_Left_2} for the earlier determined bound-state energy~$E$~(zeroing the determinant of $\mathbf{N}$), we can finally express all other wave-function amplitudes in terms of just a single one, e.g., in terms of~$ \mathcal{E} $, yielding  
    \begin{align}
        \mathcal{F} &= \Sigma_1 \, \mathcal{E} , &
        \mathcal{G} &= \Sigma_2 \, \mathcal{E} , &
        \mathcal{H} &= \Sigma_3 \, \mathcal{E} , 
    \end{align}
    and
    \begin{align}
     (\mathcal{A},\mathcal{B},\mathcal{C},\mathcal{D})&=\mathcal{E}(1,\Sigma_1,\Sigma_2,\Sigma_3)\cdot\mathbf{M}\nonumber\\
     &\equiv
     \mathcal{E}(\Lambda_1,\Lambda_2,\Lambda_3,\Lambda_4) . 
    \end{align}
    
    Employing the normalization condition for $\psi(x)$ yields a condition for $\mathcal{E}=|\mathcal{E}|$, 
    \begin{align}
        1 &= \int_{-\infty}^\infty \mathrm{d} x \, \big| \psi(x) \big|^2 \\
        &\approx \Bigg\{ \frac{\big[ | u |^2 + | v |^2 \big]}{\mathfrak{I} \big( q_{x,\mathrm{e}}^\Uparrow \big)}  \left[ \big| \Lambda_1 \big|^2 + \big| \Lambda_2 \big|^2 + 1 + \big| \Sigma_1 \big|^2 \right] \nonumber \\
        &\phantom{\Bigg\{} + \frac{\big[ | u |^2 + | v |^2 \big]}{\mathfrak{I} \big( q_{x,\mathrm{e}}^\Downarrow \big)} \left[ \big| \Lambda_3 \big|^2 + \big| \Lambda_4 \big|^2 + \big| \Sigma_2 \big|^2 + \big| \Sigma_3 \big|^2 \right] \nonumber
        \Bigg\} \big| \mathcal{E} \big|^2 ,
    \end{align}
    which we solve to consecutively determine the wave-function amplitudes and get the full Andreev-bound-state wave function $\Psi(x,y)$. 
    As a comment, each bound state with energy $\Delta_0 > E \geq 0$ comes along with a state with energy 
    $-\Delta_0 < -E \leq 0$ due to electron-hole symmetry. In what follows, we use the Andreev bound states with positive energies only.

    \subsection{Josephson current}

    Electric current through the S/F/S junction carried by Cooper pairs comes from the Andreev-reflection processes in the F~region. 
    At a given temperature~$ T $, phase difference $\varphi$, and fixed $ k_y $, the corresponding Andreev bound 
    state $ \Psi(x,y) = \psi(x) \, \mathrm{e}^{\mathrm{i} k_y y} $ contributes to the tunneling-current density an average amount of 
    \begin{align}
        j_\Psi = \lim_{x \to 0} \Bigg\{ \big\langle \psi(x>0) \big| \hat{j} \big| \psi(x>0) \big\rangle 
        \tanh \left( \frac{E}{2k_\mathrm{B}T} \right)\Bigg\}\,,
        \label{Eq_CurrentDensity} 
    \end{align}
    where 
    \begin{equation}    
        \hat{j} = -e \, \frac{\partial \hat{\mathcal{H}}_\mathrm{BdG}}{\hbar \partial k_x} = -e 
        \left[ 
        \begin{matrix} 
        1 & 0\\[0.25 cm] 
        0 & -1 
        \end{matrix} 
        \right] 
        \otimes 
        \left[ 
        \begin{matrix} 
        -\mathrm{i} \frac{\hbar}{m} \frac{\partial}{\partial x} & \mathrm{i} \frac{\alpha}{\hbar}\\[0.25 cm] 
        -\mathrm{i} \frac{\alpha}{\hbar} & -\mathrm{i} \frac{\hbar}{m} \frac{\partial}{\partial x} 
        \end{matrix} 
        \right] 
    \end{equation} 
    refers to the corresponding current-density operator along the transport direction~($\hat{x}$) that we resolved within the particle-hole space in terms of the tensor product with the $ \hat{\sigma}_z $-Pauli matrix. 
    Thereby, $ e $ stands for the (positive) elementary charge and $ k_\mathrm{B} $ for the Boltzmann constant. 
    Plugging the bound-state wave functions into the current formula, Eq.~\eqref{Eq_CurrentDensity}, and summing over all transverse channels~(i.e., integrating over all possible transverse momenta~$ k_y \in [-k_\mathrm{F};k_\mathrm{F}] $) as well as all distinct 
    bound-state branches, the total Josephson current as a function of $\varphi$~(CPR) is given by 
    \begin{widetext} 
    \begin{multline}
        I_\mathrm{J}(\varphi) = -e \sum_{E > 0} \frac{W}{2\pi} \int_{-k_\mathrm{F}}^{k_\mathrm{F}} \mathrm{d} k_y \, \frac{\hbar}{m} \Bigg\{ \mathfrak{R} \bigg[ q_{x,\mathrm{e}}^\Uparrow + \frac{\big(\alpha_\Uparrow\big)^* + \alpha_\Uparrow}{2} \lambda_\mathrm{SOC} k_\mathrm{F} \bigg] \big| \mathcal{E} \big|^2 - \mathfrak{R} \bigg[ q_{x,\mathrm{h}}^\Uparrow - \frac{\big(\tilde{\alpha}_\Uparrow\big)^* + \tilde{\alpha}_\Uparrow}{2} \lambda_\mathrm{SOC} k_\mathrm{F} \bigg] \big| \mathcal{F} \big|^2 
        \\
        + \mathfrak{R} \bigg[ q_{x,\mathrm{e}}^\Downarrow - \frac{\big(\alpha_\Downarrow\big)^* + \alpha_\Downarrow}{2} \lambda_\mathrm{SOC} k_\mathrm{F} \bigg] \big| \mathcal{G} \big|^2 - \mathfrak{R} \bigg[ q_{x,\mathrm{h}}^\Downarrow - \frac{\big(\tilde{\alpha}_\Downarrow\big)^* + \tilde{\alpha}_\Downarrow}{2} \lambda_\mathrm{SOC} k_\mathrm{F} \bigg] \big| \mathcal{H} \big|^2 \Bigg\} \big[ | u |^2 + | v |^2 \big] \tanh \left( \frac{E}{2k_\mathrm{B}T} \right) \,,
        \label{Eq_JosephsonCurrent_Final}
    \end{multline} 
    \end{widetext} 
    where $ W $ stands for the transverse width of the junction that controls the number of transverse modes~(channels) involved in transport. 
    Note that the provided Josephson-current formula is a generalization of the usual thermodynamic relation~\cite{Kulik1969,*Kulik1969alt} that computes the Josephson current from the derivatives of the Andreev-bound-state dispersion with respect to the superconducting phase difference
    $\varphi$. The advantage of Eq.~(\ref{Eq_JosephsonCurrent_Final})~is a better numerical stability and more user-friendly implementation. 
    When presenting our numerical results in~Sec.~\ref{Sec_Results}, we always normalize the Josephson current with respect to~$ ( e R_\mathrm{S} ) / (\pi \Delta_0) $, where $ R_\mathrm{S} =  ( \pi^2 \hbar ) / (e^2 k_\mathrm{F} W) $ stands for the Sharvin resistance of a perfectly transparent two-dimensional point contact; the normalized current $ ( I_{\mathrm{J}} e R_\mathrm{S} ) / (\pi \Delta_0) $ is thus independent of the transverse width~$ W $~(number of channels).
    Furthermore, the Josephson current is always evaluated at zero temperature.

    \section{Discussion of results      \label{Sec_Results}}

    To analyze the Josephson transport through the 2DEG-based S/F/S junctions, and understand the striking features of the SDE in 
    terms of spectral properties, we first calculate the Andreev-bound-state energies following the methodology sketched in~Sec.~\ref{Sec_Model}. 
    We then numerically compute the associated bound-state wave~functions and evaluate the Josephson~current by means of Eq.~\eqref{Eq_JosephsonCurrent_Final}. 
    If not otherwise stated, the $Z$-factor of the barrier equals $ Z=0.5 $, whereas the Rashba SOC~$\alpha$ and the magnetic exchange~$E_\mathrm{XC}$ are varied through their dimensionless parameters $ \lambda_\mathrm{SOC} \propto \alpha / k_\mathrm{F} $ and $ \lambda_\mathrm{XC} \propto ( E_\mathrm{XC} d ) / k_\mathrm{F} $ to scrutinize their impact on the bound-state spectra and the resulting transport properties. Our results are therefore not material-specific as they cover a wide range of junction parameters irrespective of, e.g., the Fermi energy of the 2DEG or the SOC and magnetic exchange strengths; what matters are the relative ratios with respect to the Fermi energy captured by the dimensionless~$ Z $, $ \lambda_\mathrm{SOC} $, and $ \lambda_\mathrm{XC} $.

    \subsection{Andreev~bound~states and Josephson CPRs; nonmagnetic junctions     \label{Sec_BoundStates_Nonmagnetic}}

    \begin{figure}
        \centering
        \includegraphics[width=0.49\textwidth]{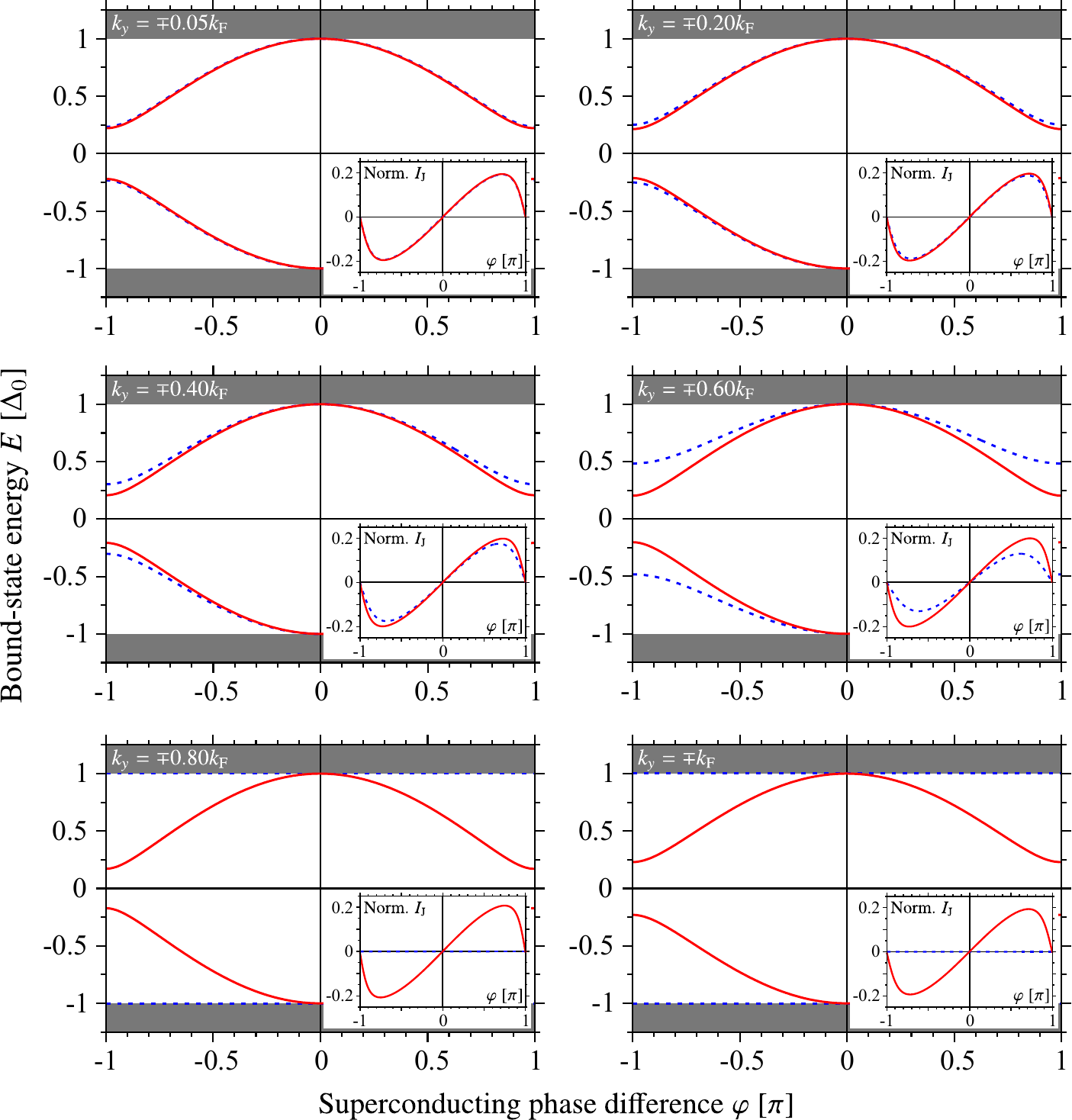}
        \caption{Calculated Andreev-bound-state energies~$ E_1^\pm $~(solid red curves) and $ E_2^\pm $~(dashed blue curves) as functions of the superconducting phase difference~$ \varphi $ for a nonmagnetic junction and several transverse momenta~$ k_y $; the other junction parameters are $ Z = 0.5 $ and $ \lambda_\mathrm{SOC} = 0.4 $. 
        The insets illustrate the $ k_y $-resolved individual contributions of the states to the total Josephson CPR. In the nonmagnetic case, the Andreev bound states~(CPRs) are symmetric~(point-symmetric) with respect to $ \varphi = 0 $. }
        \label{Fig_BoundStatesCurrent_NONMAGNETIC}
    \end{figure}

    Figure~\ref{Fig_BoundStatesCurrent_NONMAGNETIC} illustrates the dependence of the Andreev-bound-state energies on the phase difference~$ \varphi $ for nonmagnetic Josephson~junctions, $ \lambda_\mathrm{XC} = 0 $, and various transverse momenta~$ k_y $. 
    The dimensionless Rashba~parameter~$ \lambda_\mathrm{SOC} = 0.4 $ corresponds to the realistic Rashba SOC strength~\cite{Fabian2004,Fabian2007} of~$ \alpha \approx 20 \, \mathrm{meV} \, \mathrm{nm} $~(assuming a typical mass of~$ m = 0.1 m_0 $, where $ m_0 $ is the free-electron mass, and the Fermi energy of~$ \mu = 1.5 \, \mathrm{meV} $~\cite{Dartiailh2021,Lotfizadeh2023}). 

    Although the junction itself is nonmagnetic, the Rashba SOC breaks the twofold spin degeneracy of the Andreev bound states at~$k_y\neq 0$. Increasing $ |k_y| $ from~$ 0 $ to $ k_\mathrm{F} $ sweeps therefore between a scenario in which SOC does not play an important role and a regime in which the influence of SOC on the Josephson junction gets maximized. 
    For each~$ |k_y| \neq 0 $, we find two distinct spin-resolved Andreev bound states at positive energies~$ E_1^+ $~(solid red curves) and 
    $ E_2^+ > E_1^+ $~(dashed blue curves), together with their negative-energy counterparts~$ E_1^- = -E_1^+ $ and $ E_2^- = -E_2^+ $ in accordance with electron-hole~symmetry. 
    Such spin-split Andreev bound states might become relevant in connection with superconducting spin~Hall physics~\cite{Lu2009,Malshukov2010,Wang2011,Yang2012,Ren2013,Wakamura2015,Bergeret2016,Linder2017,Ouassou2017,Risinggard2019}, as they facilitate finite, transversely spin-polarized, supercurrents without the need to break the time-reversal symmetry. 
    At $ k_y = 0 $, SOC can indeed be neglected and the Andreev branches~$ E_1^\pm $ and $ E_2^\pm $ merge into single doubly spin-degenerate 
    branches, whose energies are---in good approximation---given by the well-known result valid for an effectively one-dimensional short-junction~\cite{Golubov2004,Beenakker1991,Costa2018}
    \begin{equation}
        E_1^\pm = E_2^\pm = \pm \sqrt{\frac{Z^2 + 4 \cos^2 (\varphi / 2)}{Z^2 + 4}} \Delta_0 \,.
        \label{Eq_BoundStates_Beenakker}
    \end{equation}
        
    To characterize the spin contents of the Andreev bound states, we additionally evaluated their $ \hat{\sigma}_x $-expectation values, which yield their spin projections along the current-transport axis. 
    The Andreev states with energy~$ E_1^+ $---closer to the center of the superconducting gap---have dominant 
    spin-down~(spin-up) projections along the $ \hat{x} $-axis for $ k_y>0 $~($k_y<0$), while the states with 
    energy~$ E_2^+ $---approaching the gap edge---possess dominant spin-up~(spin-down) projections for $ k_y>0 $ ($k_y<0$). The spin~features of the negative-energy states follow from time-reversal symmetry.

    The spectra of the Andreev bound states at~$ |k_y| \neq 0 $ can be explained in terms of an effective trade-off between two distinct interactions, whose strengths depend on the considered $k_y$-channel. 
    On the one hand, if SOC is completely absent, the energies of the Andreev bound states can be extracted from~Eq.~\eqref{Eq_BoundStates_Beenakker} replacing $ Z $ by 
    the effective, channel-dependent, barrier~parameter~$ Z[k_y]=Z / \sqrt{1 - k_y^2 / k_\mathrm{F}^2} $~\cite{Costa2018}.
    This substitution accounts for the reduced wave-vector~projection~(electron velocity) along the $ \hat{x} $-current direction with increasing~$ |k_y| $, which makes it basically harder for electrons to tunnel through the junction in channels with large~$ |k_y| $. 
    Increasing~$ |k_y| $ corresponds therefore to introducing larger, and channel-dependent, effective barrier~strengths~$ Z[k_y] $ and expels the Andreev states~(independent of their spin) spectrally from the center of the gap, reaching~$ E_1^\pm = E_2^\pm \approx \pm \Delta_0 $ in the limit of~$ |k_y| \to k_\mathrm{F} $. 
    On the other hand, nonzero SOC rises the additional potential~$ \langle \hat{\mathcal{H}}_\mathrm{SOC} \rangle[k_y] \approx \alpha k_y \langle \hat{\sigma}_x \rangle $ in the Hamiltonian in~Eq.~\eqref{Eq_BdG_2}. As the sign of $ k_y\langle \hat{\sigma}_x \rangle $ differs for the $ E_1^\pm $- and $ E_2^\pm $-states~(due to their opposite $ \langle \hat{\sigma}_x \rangle $-spin contents mentioned above), this term acts as a spin-split potential that effectively subtracts from or adds to the global spin-independent tunnel potential $V$~(or its dimensionless counterpart $Z$). As a result, the scattering for 
    the $ E_1^\pm $-branches gets suppressed, while that for the $ E_2^\pm $-branches is enhanced. 
    Thus, increasing~$ |k_y| $ spectrally pushes the $ E_1^\pm $-states more toward the center of the superconducting gap, while the $ E_2^\pm $-states are expelled toward the gap~edges, as shown in~Fig.~\ref{Fig_BoundStatesCurrent_NONMAGNETIC}~(see, for~instance, the panel for $ |k_y| = 0.60 k_\mathrm{F} $).

    Inspecting~Eq.~\eqref{Eq_WV_SpinUp} for $\lambda_{\mathrm{SOC}}>0$, we see that there exists a critical~magnitude of the transverse momentum 
    \begin{equation}
        |k_y^\mathrm{crit.}| = k_\mathrm{F} \left( \sqrt{1 + \lambda_\mathrm{SOC}^2} - \lambda_\mathrm{SOC} \right) ,
        \label{Eq_Critky}
    \end{equation}
    such that the spin-up wave vectors $q_{x,\mathrm{e}}^\Uparrow$ and $ q_{x,\mathrm{h}}^\Uparrow$ of quasiparticles become purely imaginary for $ |k_y|>|k_y^\mathrm{crit.}| $, and the corresponding $ k_y $-modes no longer contribute to the Josephson current---for $\lambda_\mathrm{SOC}=0.4$, $|k_y^\mathrm{crit.}| \approx 0.68 k_\mathrm{F}$. 
    In terms of Andreev bound states, the $ E_2^\pm $-branches at~$ |k_y| > |k_y^\mathrm{crit.}| $ become indeed flat bands~(with respect to~$ \varphi $) pinned at the gap-edges~$ \pm \Delta_0 $ that do not carry any Josephson currents, as seen, e.g., for $ |k_y| = 0.80 k_\mathrm{F} $ in~Fig.~\ref{Fig_BoundStatesCurrent_NONMAGNETIC}.     
    These arguments do not apply to the spin-down quasiparticle wave vectors, Eq.~\eqref{Eq_WV_SpinDown}, which remain real for all $\lambda_{\mathrm{SOC}}>0$, and allow 
    for the current-carrying in-gap $ E_1^\pm $-states at any~$ |k_y|<k_\mathrm{F} $, establishing the $ E_1^\pm $-Andreev states as the main channels to transport Cooper pairs.

    \begin{figure}
        \centering
        \includegraphics[width=0.49\textwidth]{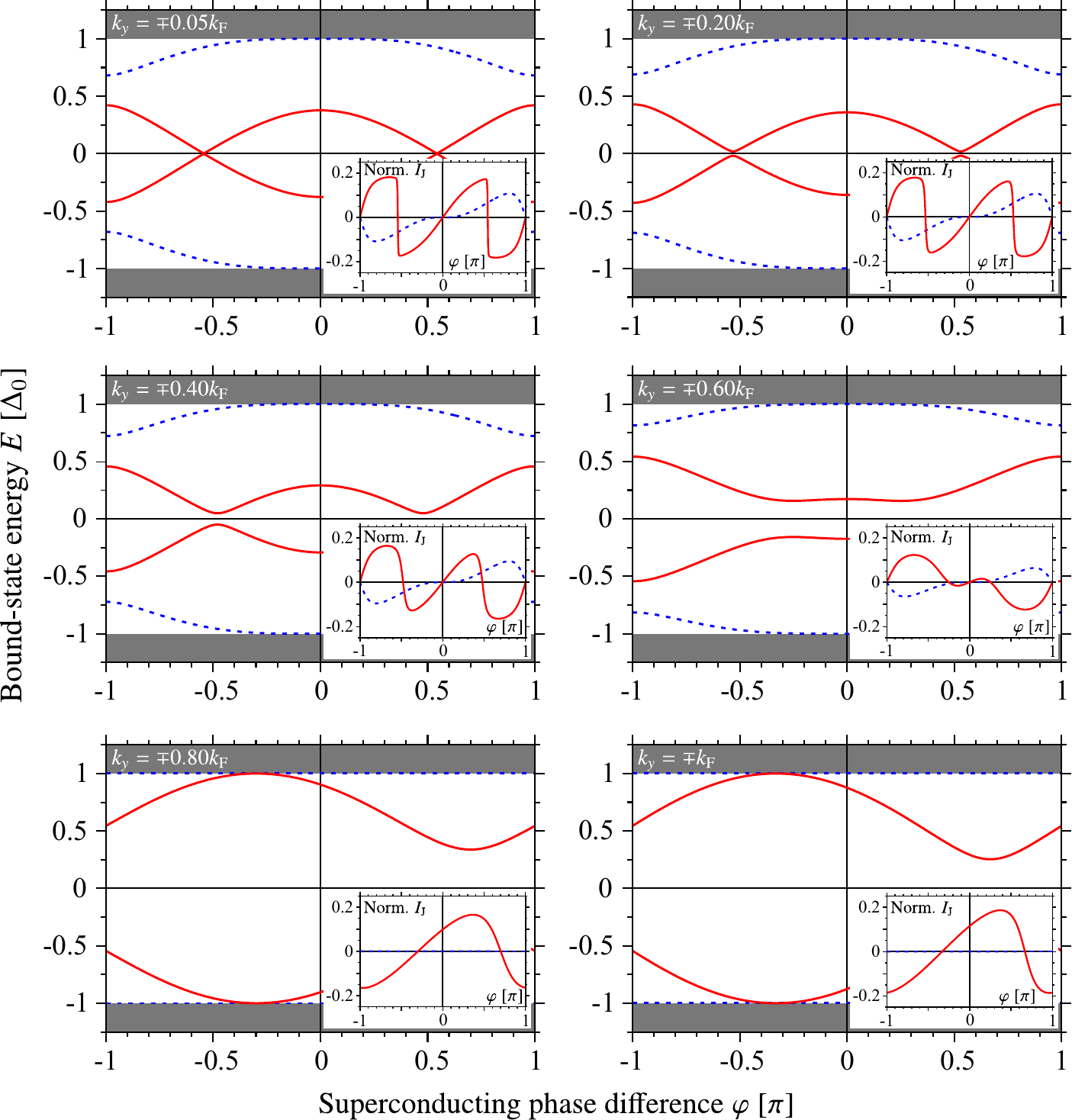}
        \caption{Calculated Andreev-bound-state energies~$ E_1^\pm $~(solid red curves) and $ E_2^\pm $~(dashed blue curves) as functions of the superconducting phase difference~$ \varphi $ for a magnetic junction and several transverse momenta~$ k_y $; the junction parameters are $ Z = 0.5 $, $ \lambda_\mathrm{SOC} = 0.4 $, and $ \lambda_\mathrm{XC} = 1.5 $. 
        The insets illustrate the $ k_y $-resolved individual contributions of the states to the total Josephson CPR. In the magnetic case, the Andreev bound states and CPRs develop an asymmetry with respect to $ \varphi = 0 $ at larger $|k_y|$~(i.e., at~$ |k_y| \gtrsim |k_y^\mathrm{crit.}| $). When superimposed, the different asymmetries of different $ k_y $-channels imprint a global $ \varphi_0 $-distortion on the total Josephson CPR and result in the Josephson SDE. The junction is overall in the $0$-like state. }
        \label{Fig_BoundStatesCurrent_MAGNETIC}
    \end{figure}

    \begin{figure}
        \centering
        \includegraphics[width=0.49\textwidth]{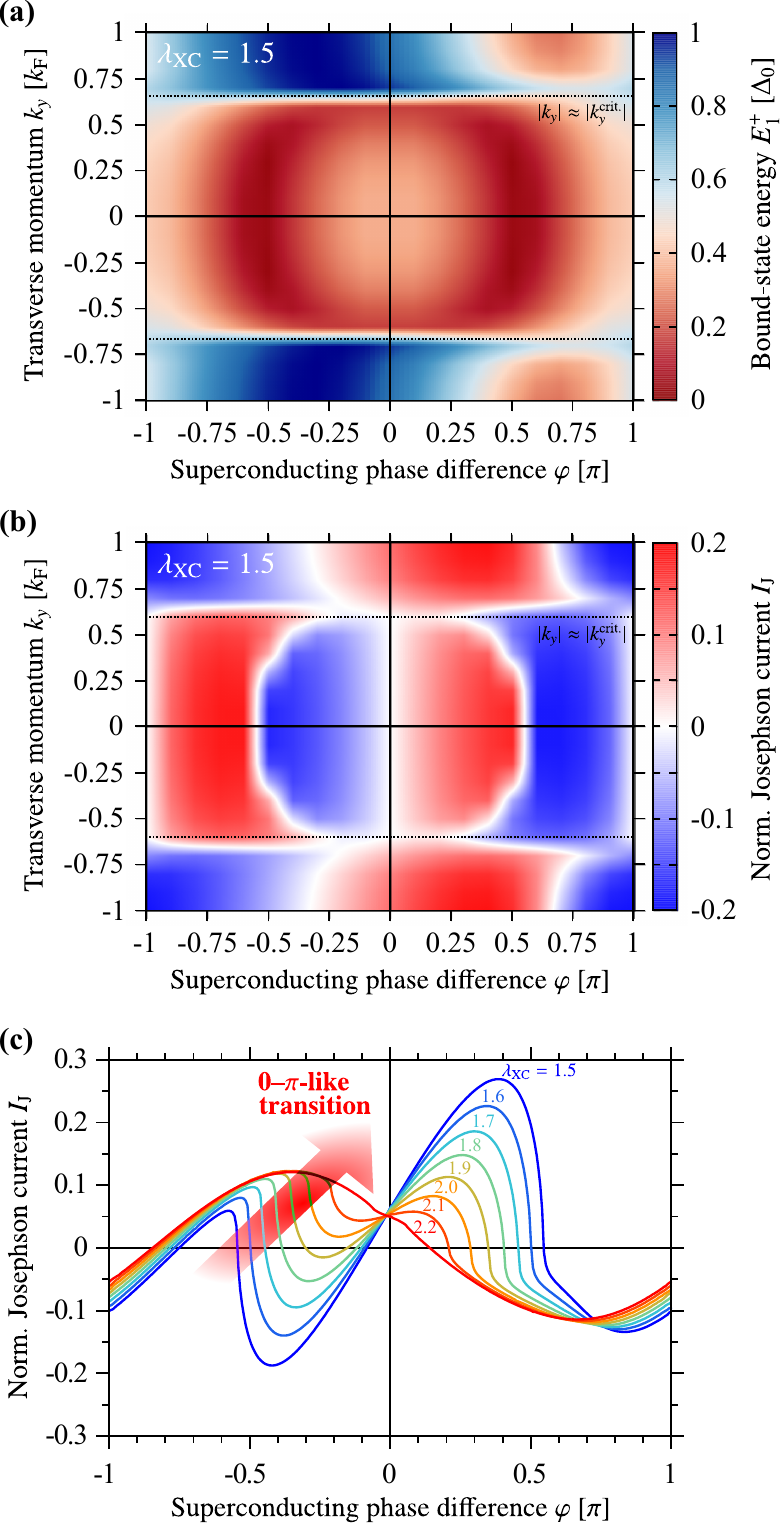}
        \caption{(a)~Andreev-bound-state energies and (b)~individual Josephson CPRs of the transport-dominating $ E_1^+ $-Andreev branch, displayed as color maps with a dense sampling of $ \varphi $ and~$ k_y $; the junction parameters are $ Z = 0.5 $, $ \lambda_\mathrm{SOC} = 0.4 $, and $ \lambda_\mathrm{XC} = 1.5 $. 
        When the transverse momentum $|k_y|$ exceeds its critical $|k_y^\mathrm{crit.}| $~[defined by~Eq.~(\ref{Eq_Critky})], both quantities show a clear asymmetry in $\varphi$---acting as a precursor of strongly distorted total Josephson CPRs and the resulting Josephson SDE. 
        (c)~Total Josephson CPRs, summed over all transverse $ k_y $-channels, for the indicated values of 
        the magnetic exchange~$ \lambda_\mathrm{XC} $; the remaining junction parameters are the same as in panels~(a) and~(b). 
        Apart from the pronounced Josephson SDE---i.e., clearly different magnitudes of positive and negative critical currents---the junction undergoes a $ 0 $--$ \pi $-like transition when raising $\lambda_\mathrm{XC}$ from 1.5 to 2.2. }
        \label{Fig_BoundStatesCurrent_COLORMAP}
    \end{figure}

    \subsection{Andreev~bound~states and Josephson CPRs; magnetic junctions     \label{Sec_BoundStates_Magnetic}}

    S/F/S junctions with simultaneously broken space-inversion and time-reversal symmetries were intensively studied~\cite{Bezuglyi2002,Krive2004,Buzdin2008,Reynoso2008,Zazunov2009,Liu2010a,Liu2010,Liu2011,Reynoso2012,YokoyamaJPSJ2013,Brunetti2013,Shen2014,Yokoyama2014Anomalous,Konschelle2015,Szombati2016,Assouline2019,Mayer2020b,Strambini2020} in connection with the anomalous Josephson effect leading to $ \varphi_0 $-shifted CPRs, provided that the exchange field has a nonzero component perpendicular to the current flow. 
    In the magnetic 2DEG-based Josephson junction with Rashba SOC considered in this work, it is crucial that many $ k_y $-channels contribute to transport. As we will demonstrate below, the CPRs of different $ k_y $-channels will acquire distinct individual $ \varphi_0 $-shifts that depend on~$ |k_y| $ and become maximal as~$ |k_y| \to k_\mathrm{F} $. 
    The superposition of all $ k_y $-channels' CPRs gives the total CPR~$I_\mathrm{J}(\varphi)$, Eq.~\eqref{Eq_JosephsonCurrent_Final}, then an anharmonically distorted shape with a global $\varphi_0$-shift, and with different magnitudes of positive---$ I_\mathrm{c}^+ $---and negative---$ I_\mathrm{c}^- $---critical currents. Such junctions can therefore act as \emph{Josephson supercurrent diodes}, and polarity-wise rectify supercurrents from dissipative currents.

    Figure~\ref{Fig_BoundStatesCurrent_MAGNETIC} shows the Andreev spectra 
    for magnetic Josephson~junctions with the same values of $Z$ and $\lambda_\mathrm{SOC}$, and
    the same transverse momenta~$ k_y $, as in the nonmagnetic case. 
    For the exchange parameter, we use~$ \lambda_\mathrm{XC} = 1.5 $, which becomes equivalent to $ E_\mathrm{XC} d \approx 36 \, \mathrm{meV} \, \mathrm{nm} $~(assuming the same mass and Fermi energy as stated in the nonmagnetic part); for a typical link thickness of about~$ d = 100 \, \mathrm{nm} $~\cite{Baumgartner2021,Baumgartner2022a,Baumgartner2022,Costa2023,*Costa2023NatNano}, this gives feasible values of the exchange splitting below $ 1 \, \mathrm{meV} $ that could be experimentally realized either through proximity to a strong ferromagnetic top layer or through the Zeeman splitting induced by magnetic fields of a few hundred~$ \mathrm{mT} $~(depending on the $ g $-factor of the 2DEG).

    The numerically obtained bound-state spectral characteristics can be qualitatively understood in a similar way as in the nonmagnetic case, accounting for the additional exchange-generated potential $\langle \hat{\mathcal{H}}_\mathrm{XC} \rangle[k_y]$ that now needs to be considered along with $Z[k_y]$ and $ \langle \hat{\mathcal{H}}_\mathrm{SOC} \rangle[k_y]$. 
    In what follows, we summarize the main findings of our numerical analyses. The exchange term expels the $ E_2^\pm $-states even stronger from the center of the superconducting gap than in the nonmagnetic case, indicating that the $E_2^\pm$-branches play again only a very minor role for the total Josephson current. Contrary, the $ E_1^\pm $-states remain well inside the superconducting gap, carry most of the Josephson current, and can---at certain exchange splittings~$ \lambda_\mathrm{XC} $ that depend on $k_y$---even cross zero~energy; see Fig.~\ref{Fig_BoundStatesCurrent_MAGNETIC}. 
    Computing the individual CPRs associated with the $ E_1^\pm $-states of the distinct $ k_y $-channels~(see the insets in~Fig.~\ref{Fig_BoundStatesCurrent_MAGNETIC}), we deduce that these zero-energy crossings of the $ E_1^\pm $-states signify the exchange splittings at which the CPRs of these $ k_y $-channels experience $ 0 $--$ \pi $~transitions~\cite{Bulaevskii1977a,*Bulaevskii1977b,Ryazanov2001,Andersen2006,Kawabata2010,Kawabata2012} and their current contributions reverse their sign~(at a fixed phase difference~$ \varphi $). As the total Josephson current is nevertheless still the superposition of many independent transverse channels, $ 0 $--$ \pi $~transitions in single $ k_y $-channels do usually not yet facilitate a global $ 0 $--$ \pi $~transition of the whole 2DEG-based junction~\cite{Fang2023}. An experimental study of the transition regime in two-dimensional junctions could hence become quite challenging, as one would need to individually address transport through single $ k_y $-channels to detect the $ 0 $--$ \pi $~transitions.

    Zero-energy (bound) states, accompanied by changes of the particle-hole, spatial, and spin contents~\cite{Costa2018} of the ground-state wave function, were already proposed to be indicative of phase transitions back in the 1970s~\cite{Sakurai1970}. We applied this concept to one-dimensional S/F/S Josephson junctions in our earlier work~\cite{Costa2018}, and showed that this phase transition is indeed the current-reversing $ 0 $--$ \pi $~transition. 
    \footnote{Plotting the bound-state spectrum versus the magnetic exchange~parameter~$ \lambda_\mathrm{XC} $ reveals that the $ E_1^\pm $-bands follow the dispersion of the well-known Yu--Shiba--Rusinov~states~\cite{Yu1965,Shiba1968,Shiba1969,Rusinov1968,*Rusinov1968alt}. For this reason, we termed these states \emph{Yu--Shiba--Rusinov-like states} in~Ref.~\cite{Costa2018} to distinguish them from the Andreev-like $ E_2^\pm $-states. However, as this is not relevant to the studied physics, we use the term Andreev bound states for $ E_1^\pm $ \emph{and} $ E_2^\pm $ in this work. }. 
    Compared to the one-dimensional case, it is worth mentioning that the strength of~$ \lambda_\mathrm{XC} $ that is necessary to trigger the zero-energy crossings of the Andreev bound states in the 2DEG-based junction additionally depends on $k_y$. Specifically, while the~$ k_y = 0 $-channel requires the largest exchange, the channels with elevated~$ |k_y| $ already allow for zero-energy Andreev states at (much) smaller exchange. 
    If the Rashba SOC is weak compared to the exchange, the bound-state energies of one-dimensional S/F/S junctions are well describable by an analytical formula that is at full length stated as~Eq.~(2) in~Ref.~\cite{Costa2018}. 
    To generalize this formula to the present two-dimensional context, one needs to resubstitute
    $Z$~($ \overline{\lambda}_\mathrm{SC} $ in~Ref.~\cite{Costa2018}) and $ \lambda_\mathrm{XC} $~($ \overline{\lambda}_\mathrm{MA} $ in~Ref.~\cite{Costa2018}) for each individual $ k_y $-channel as 
    \begin{equation}
        Z[k_y]=\frac{Z}{\sqrt{1 - k_y^2 / k_\mathrm{F}^2}} 
        \ \ \text{and}\ \ 
        \lambda_\mathrm{XC} [k_y]=\frac{\lambda_\mathrm{XC}}{\sqrt{1 - k_y^2 / k_\mathrm{F}^2}}\,.
    \end{equation}
    As we briefly mentioned in the nonmagnetic case, this resubstitution effectively accounts for the lowered $ \hat{x} $-projections of the electron velocities due to their nonzero transverse momenta in channels with~$ |k_y| \neq 0 $. 
    Thus, the Andreev bound states forming in the two-dimensional magnetic junction~(and at weak SOC) behave even for $ |k_y| \neq 0 $ still similarly to their one-dimensional counterparts, but with the ``renormalized'' channel-specific $Z[k_y]$ and $ \lambda_\mathrm{XC}[k_y] $ barrier and exchange parameters, respectively. This observation has a profound impact since even an exchange~$ \lambda_\mathrm{XC} $ that is too weak to induce $ 0 $--$ \pi $~transitions in the one-dimensional case can now cause a ``renormalized'' $ \lambda_\mathrm{XC}[k_y] $ at large enough~$ |k_y| $ that is sufficient to trigger such transitions---at least in the individual CPRs of these $ k_y $-channels; see, e.g., our numerical calculations in~Fig.~\ref{Fig_BoundStatesCurrent_MAGNETIC}.

    \begin{figure}
        \centering
        \includegraphics[width=0.49\textwidth]{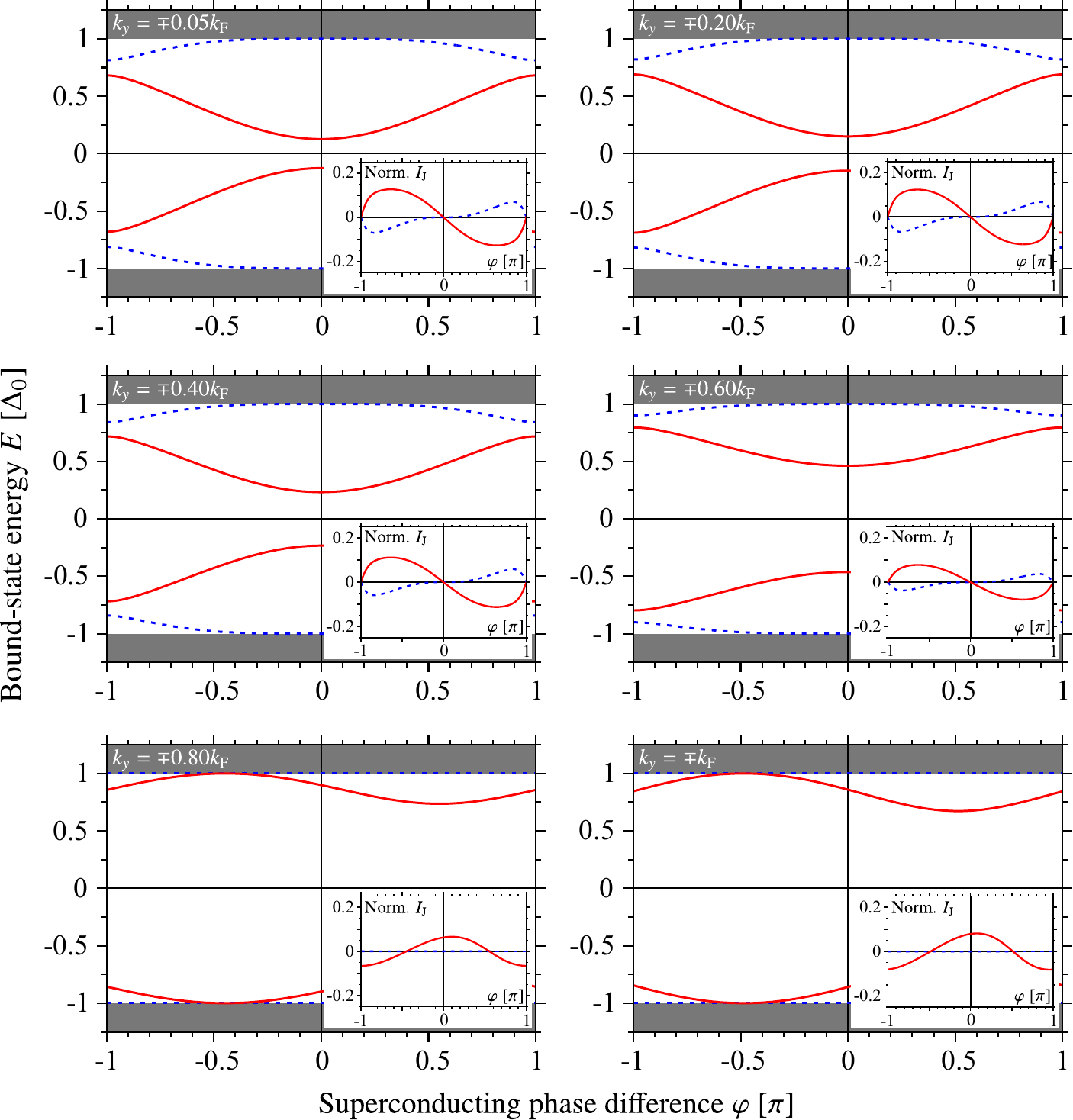}
        \caption{Calculated Andreev-bound-state energies~$ E_1^\pm $~(solid red curves) and $ E_2^\pm $~(dashed blue curves) as functions of the superconducting phase difference~$ \varphi $ for a strongly magnetic junction and several transverse momenta~$ k_y $; the junction parameters are $ Z = 0.5 $, $ \lambda_\mathrm{SOC} = 0.4 $, and $ \lambda_\mathrm{XC} = 2.5 $. 
        The insets illustrate the $ k_y $-resolved individual contributions of the states to the total Josephson CPR. The junction is overall in the $\pi$-like state.} 
        \label{Fig_BoundStatesCurrent_STRONGLY_MAGNETIC}
    \end{figure}

    Analogously to the nonmagnetic scenario, the Andreev-bound-state spectra for $ |k_y| > |k_y^\mathrm{crit.}| $ support only 
    current-carrying $ E_1^\pm $-in-gap states, while their $ E_2^\pm $-counterparts are pinned to the gap edges~$\pm \Delta_0 $ and again flat bands~(with respect to~$ \varphi $) not contributing to the Josephson current. Remarkably,
    for $|k_y| > |k_y^\mathrm{crit.}| $~($ \approx 0.68 k_\mathrm{F} $ at $\lambda_\mathrm{SOC}=0.4$), 
    (i)~the $ E_1^\pm $-bands manifestly lack the symmetry with respect to a reversal of $\varphi$ and (ii)~the CPRs, associated with the $ E_1^\pm $-bound states, acquire clearly observable and \emph{for the individual $ k_y $-channels different in magnitude $\varphi_0$-shifts}~(whose amplitudes continuously increase with increasing~$ |k_y| $), which---after averaging over all channels' CPRs~\cite{Costa2023,*Costa2023NatNano} 
    \footnote{
    The low-temperature Josephson CPRs of the individual $ k_y $-channels of a short ballistic and highly-transparent junction can, in general, be expanded into a series of (higher-harmonic) sinusoidal contributions
    \begin{equation*}
        I_\mathrm{J} (\varphi ; k_y) = a_1 (k_y) \sin(\varphi) + a_2 (k_y) \sin(2\varphi) + \ldots ,
    \end{equation*}
    where the expansion coefficients depend on the considered $ k_y $-channel. 
    The simultaneous breaking of space-inversion and time-reversal symmetries additionally superimposes $ k_y $-dependent (and, in general, different due to the $ k_y $-dependent Fermi-velocity projection along $ \hat{x} $~\cite{Buzdin2008}) $ \varphi_0 $-shifts on each higher-harmonic term, i.e., 
    \begin{align*}
        I_\mathrm{J} (\varphi ; k_y) &= a_1 (k_y) \sin \big( \varphi + \varphi_{0}^{(1)} (k_y) \big) \\
        &\hspace{25 pt} + a_2 (k_y) \sin \big( 2\varphi + \varphi_0^{(2)} (k_y) \big) + \ldots , 
    \end{align*}
    where $ \varphi_0^{(i)} (k_y) $ indicates the $ \varphi_0 $-phase shift of the $ i $th higher-harmonic CPR term in the regarded $ k_y $-channel. 
    Direction-dependent critical currents~(and hence the SDE) occur in a \emph{single} $ k_y $-channel only if $ \varphi_0^{(n)} \neq n \varphi_0^{(1)} $; otherwise, the only effect on the $ k_y $-resolved CPRs is a horizontal shift~(along the $ \varphi $-axis), as we observe, e.g., in~Fig.~\ref{Fig_BoundStatesCurrent_MAGNETIC}. 
    However, in the presence of many distinct $ k_y $-channels, the \emph{total} Josephson CPR~$ I_\mathrm{J} (\varphi) $ is obtained summing over all channel-resolved CPRs, 
    \begin{equation*}
        I_\mathrm{J} (\varphi) = \sum_{k_y} I_\mathrm{J} (\varphi ; k_y) ;
    \end{equation*}
    as the $ \varphi_0^{(i)} (k_y) $ are now generally different in different $ k_y $-channels, the individual phase shifts will never fully compensate, and the total CPR experiences the characteristic distortion~(global $ \varphi_0 $-shift) illustrated in~Fig.~\ref{Fig_BoundStatesCurrent_COLORMAP}(c) with direction-dependent critical currents~(and hence the SDE).}
    ---results in a nonzero global $ \varphi_0 $-shift and distortion of the \emph{total} CPR with different magnitudes of the positive and negative critical Josephson currents, and thereby in the Josephson SDE; see Figs.~\ref{Fig_BoundStatesCurrent_MAGNETIC} and \ref{Fig_BoundStatesCurrent_COLORMAP}(c).  

    Why do we not observe $ \varphi_0 $-shifts at $ |k_y| < |k_y^\mathrm{crit.}| $? As we explained earlier, it is the simultaneous breaking of space-inversion and time-reversal symmetries that causes the anomalous $ \varphi_0 $~(and finally also the SDE), suggesting that actually all transverse channels should develop finite $ \varphi_0 $-phase shifts in their CPRs, independent of their respective $ |k_y| $. However, this would require that the magnetic link had a finite thickness~(like in real junctions) such that solving the Bogoliubov--de Gennes equation yields Andreev-state wave functions that inherently break time-reversal symmetry through the exchange-split wave vectors inside the link. To provide an analytical study, which allows us to relate the SDE characteristics to spectral properties of Andreev bound states, we reduced the magnetic link to the delta-like form, where the breakdown of the time-reversal symmetry enters through the interfacial boundary conditions~[see~Eq.~\eqref{Eq_Boundary_Right}]. The time-reversal-symmetry breaking is therefore not explicitly seen for all $k_y$-modes, but only if $|k_y|$ exceeds a certain threshold; this turns out to be equivalent to~$ |k_y^\mathrm{crit.}| $ defined by~Eq.~\eqref{Eq_Critky}. Then, the spin-up and spin-down components of the Andreev-state wave functions~$ \Psi(x,y) $~[recall Eq.~\eqref{Eq_WaveFunction}] acquire different spectral characters---the first is described by an evanescent wave, as the spin-up wave vector becomes fully imaginary, while the second still corresponds to propagating solutions---maximally breaking the spin~(time-reversal) symmetry of the Andreev states and leading to sizable $ \varphi_0 $. 

    The existence of the threshold~$ |k_y^\mathrm{crit.}| $ and formation of $ \varphi_0 $ only in channels with~$ |k_y| \gtrsim |k_y^\mathrm{crit.}| $ are hence specific to the applied delta-like model, and the model might initially seem to be quite a rigid approximation of real junctions. Nevertheless, a detailed analysis of the results extracted from the delta-like model and comparison with the experimental data of Al-gated InAs 2DEG-based Josephson junctions~\cite{Costa2023,*Costa2023NatNano} in the short-junction limit~\cite{Scharf2019,Baumgartner2021,Lotfizadeh2023}, in which the thickness of the link is small compared to the superconducting coherence length, showed that our approach is valid in this regime to capture the fundamental SDE physics. The $ \varphi_0 $-shifts are, in general, underestimated in our model as the contributions of the $ |k_y| < |k_y^\mathrm{crit.}| $-channels are not included. This could be compensated through a rescaling of the phenomenological magnetic-exchange parameter~$ \lambda_\mathrm{XC} $, as we demonstrate in~Ref.~\cite{Costa2023,*Costa2023NatNano}.

    So far, we displayed the Andreev bound states and their contributions to the Josephson current for a discrete set of transverse momenta $|k_y|$---namely, the~0.05, 0.20, 0.40, 0.60, 0.80, and 1 multiples of the Fermi wave vector~$k_\mathrm{F}$. 
    To convincingly show all the trends on a finer scale, Figs.~\ref{Fig_BoundStatesCurrent_COLORMAP}(a) and \ref{Fig_BoundStatesCurrent_COLORMAP}(b) illustrate the Andreev-bound-state energies and their Josephson CPRs 
    as color plots with a dense sampling of the~$ \varphi $- and $ k_y $-ranges; all other parameters are the same as before and, for simplicity, we just 
    show the transport-dominating $ E_1^\pm $-states. 
    The numerical results fully confirm our previous conclusions, now at generic $ |k_y| $: (i)~the zero-energy crossings 
    of the $ E_1^\pm $-states coincide with $ 0 $--$ \pi $~reversals of the Josephson current of the corresponding $ k_y $-channels and (ii)~the strong asymmetry with respect to $ \varphi = 0 $, coming along with pronounced $ \varphi_0 $-shifts, clearly develops for the bound states of the channels with~$ |k_y| > |k_y^\mathrm{crit.}| $. 
    As the total Josephson CPR of the junction with given~$Z$, $\lambda_\mathrm{SOC}$, and $ \lambda_\mathrm{XC} $ comes as a sum of all its transverse channels, Eq.~(\ref{Eq_JosephsonCurrent_Final}), the $\varphi_0$-shifts developed in the individual $ k_y $-channels give the total Josephson current a pronounced anharmonic $ \varphi $-dependence. 
    
\begin{figure}
        \centering
        \includegraphics[width=0.49\textwidth]{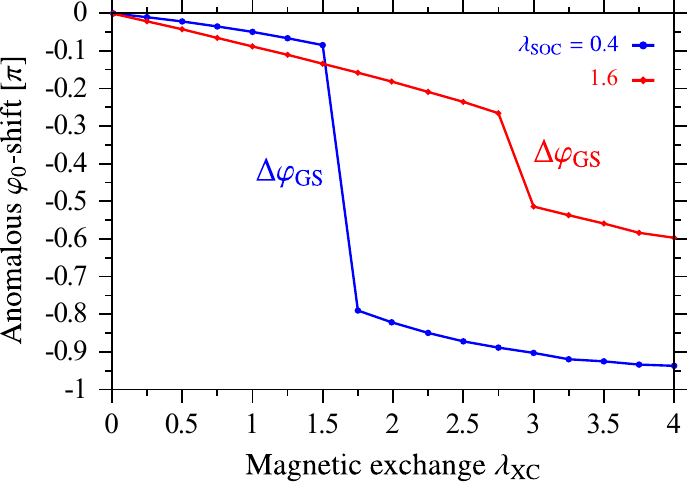}
        \caption{Anomalous (global) $ \varphi_0 $-phase shifts extracted from the total Josephson CPRs $I_\mathrm{J}(\varphi)$ as functions of the magnetic exchange~$ \lambda_\mathrm{XC} $ and for two different Rashba-SOC parameters~$ \lambda_\mathrm{SOC} $, keeping $Z=0.5$ constant. The jumps in $\varphi_0$, indicated as $ \Delta \varphi_\mathrm{GS} $, correspond to the jumps of the ground-state phases that the junction experiences when undergoing  $0$--$\pi$-like transitions~[the blue curve refers to the $ 0 $--$ \pi $-like transition displayed in~Fig.~\ref{Fig_BoundStatesCurrent_COLORMAP}(c)]. }
            \label{Fig_Phi0}
    \end{figure}

    For the sake of completeness, and to better visualize the aforementioned anharmonicity, we present the full CPRs of a junction close to the $ 0 $--$ \pi $~transition, varying~$ \lambda_\mathrm{XC} $ in steps of 0.1 from 1.5 to 2.2, in~Fig.~\ref{Fig_BoundStatesCurrent_COLORMAP}(c). For $\lambda_\mathrm{XC} \lesssim 1.7 $, the junction is in a $0$-like state~(see Fig.~\ref{Fig_BoundStatesCurrent_MAGNETIC} for the spectral characteristics), while it is already in a $ \pi $-like state at $ \lambda_\mathrm{XC} \approx 2.0 $~(see Fig.~\ref{Fig_BoundStatesCurrent_STRONGLY_MAGNETIC} for the spectral characteristics); the interim exchange range between 1.7 and 2.0 spans the transition region in which the state of the junction is usually a mixture of both and hard to uniquely discern. The reason to term the states $ 0 $- or $ \pi $-\emph{like} is as follows~\cite{YokoyamaJPSJ2013,Yokoyama2014Anomalous}. Given the CPR~$I_\mathrm{J}(\varphi)$, we compute the Josephson energy~$ E_\mathrm{J}(\varphi)\propto\int_0^\varphi \mathrm{d} \phi \, I_\mathrm{J}(\phi) $ as a function of the superconducting phase difference~$ \varphi $ and look for the position of its global minimum to determine the energetically favored junction ground state. The minima are neither strictly at $0$ nor at $\pi$ phase difference, but close to these values motivating to call the corresponding junction states $ 0 $-\emph{like} in the first and $ \pi $-\emph{like} in the second case. For the same reason, the phase jump of the ground-state wave function when the junction undergoes the $ 0 $--$ \pi $~transition is also not strictly $ \pi $, but instead rather a generic $ |\Delta \varphi_\mathrm{GS}| \neq \pi $, and we likewise term the transition $ 0 $--$ \pi $-\emph{like}. 
    The $ \Delta \varphi_\mathrm{GS} $-phase jumps during the $ 0 $--$ \pi $-like transitions are illustrated for two different SOCs in the plot of the global $\varphi_0$-shifts of the total Josephson CPRs versus $\lambda_\mathrm{XC}$ that we provide in~Fig.~\ref{Fig_Phi0}.

    \subsection{Characterization of the Josephson SDE     \label{Sec_Diode}}

    \begin{figure}
        \centering
        \includegraphics[width=0.49\textwidth]{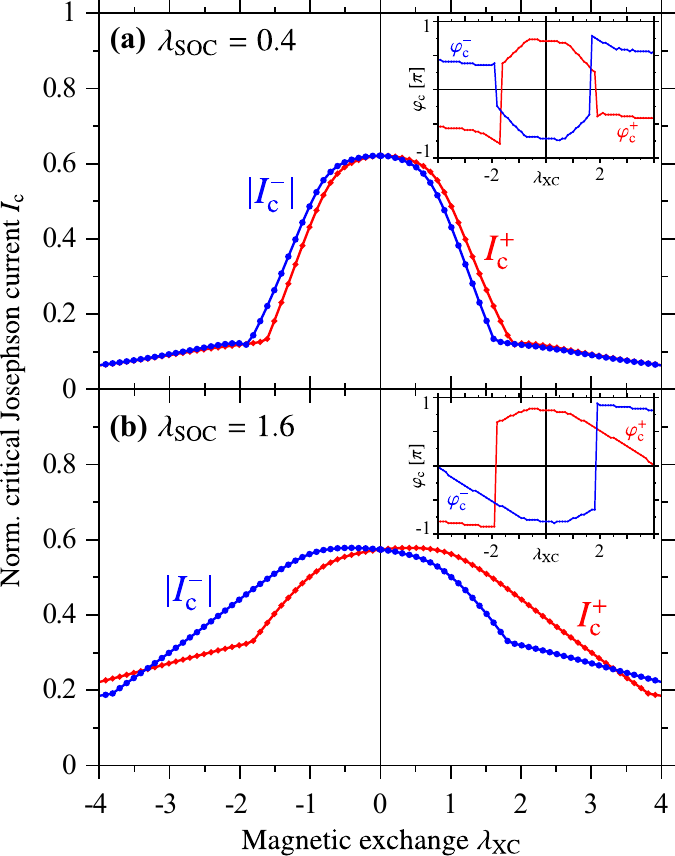}
        \caption{Polarity dependence of the critical Josephson currents~$ I_\mathrm{c}^+ $ and $ |I_\mathrm{c}^-| $, displayed as functions of the magnetic exchange~$ \lambda_\mathrm{XC} $; the other junction parameters are (a)~$ Z = 0.5 $ and $ \lambda_\mathrm{SOC} = 0.4 $, as well as (b)~$ Z = 0.5 $ and $ \lambda_\mathrm{SOC} = 1.6 $. 
        The insets show the modulations of the critical superconducting phase differences $ \varphi_\mathrm{c}^+ $ and $ \varphi_\mathrm{c}^- $---defined by $ I_\mathrm{c}^+ = I_\mathrm{J} (\varphi_\mathrm{c}^+) $ and $ I_\mathrm{c}^- = I_\mathrm{J} (\varphi_\mathrm{c}^-) $---with~$ \lambda_\mathrm{XC} $. Jumps in $ \varphi_\mathrm{c}^\pm $ correspond to cusps~(smoothness fractures) in $ I_\mathrm{c}^+ $ and $ |I_\mathrm{c}^-| $, and indicate the ``first'' and ``second acts'' of the $ 0 $--$ \pi $-like transition, as described in the text.}
        \label{Fig_CritCurrent_DifferentSOC_ReducedSize}
    \end{figure}

    To further explore the Josephson SDE, we compute the Josephson~CPRs at different values of magnetic exchange~$ \lambda_\mathrm{XC} $, and extract the corresponding positive and negative critical currents~$ I_\mathrm{c}^+ $ and 
    $ I_\mathrm{c}^- $---serving as the global maxima and minima of $I_\mathrm{J}(\varphi)$ along the interval~$ \varphi \in [-\pi ; \pi] $. The results are illustrated 
    in~Figs.~\ref{Fig_CritCurrent_DifferentSOC_ReducedSize}(a) and \ref{Fig_CritCurrent_DifferentSOC_ReducedSize}(b) for the Rashba SOCs~$ \lambda_\mathrm{SOC} = 0.4 $ and $ \lambda_\mathrm{SOC} = 1.6 $, respectively. 
    
    In the nonmagnetic junction~(i.e., when~$ \lambda_\mathrm{XC} = 0 $), the total CPR~$I_\mathrm{J}(\varphi)$ is always strictly point-symmetric with respect to zero phase~(recall~Fig.~\ref{Fig_BoundStatesCurrent_NONMAGNETIC}), and both the positive and negative critical currents~$ I_\mathrm{c}^+ $ and $ I_\mathrm{c}^- $ are equal in magnitude. 
    Increasing the magnetic exchange~$ \lambda_\mathrm{XC} $ distorts $I_\mathrm{J}(\varphi)$, as we explained in detail in~Sec.~\ref{Sec_BoundStates_Magnetic}, and produces a noticeable difference between $ I_\mathrm{c}^+ $ and $ |I_\mathrm{c}^-| $, i.e., the Josephson SDE. 
    Enlarging~$ \lambda_\mathrm{XC} > 0 $ suppresses $ |I_\mathrm{c}^-| $ stronger than~$ I_\mathrm{c}^+ $, particularly at an elevated value of SOC, and vice versa for~$ \lambda_\mathrm{XC} < 0 $. 
    Scrutinizing Figs.~\ref{Fig_CritCurrent_DifferentSOC_ReducedSize}(a) and \ref{Fig_CritCurrent_DifferentSOC_ReducedSize}(b) in greater detail, we observe that the $ |I_\mathrm{c}^-| $-curves develop local cusps, beyond which the amplitudes of 
    $ |I_\mathrm{c}^-| $ decrease slower with further increasing~$ \lambda_\mathrm{XC} $, at~$ \lambda_\mathrm{XC} \approx 1.5 $ for  $ \lambda_\mathrm{SOC} = 0.4 $ and $ \lambda_\mathrm{XC} \approx 1.8 $ for $ \lambda_\mathrm{SOC} = 1.6 $. 
    To unravel the physical meaning of these cusps, the insets of~Fig.~\ref{Fig_CritCurrent_DifferentSOC_ReducedSize} illustrate the $\lambda_\mathrm{XC} $-dependence of the critical superconducting phase differences~$ \varphi_\mathrm{c}^+ $ and $ \varphi_\mathrm{c}^- $, which are the phase differences corresponding to the critical currents~$ I_\mathrm{c}^+ $ and $ I_\mathrm{c}^- $. 
    The cusps in $ |I_\mathrm{c}^-| $ indeed correspond to those $ \lambda_\mathrm{XC} $ at which the critical phase~$ \varphi_\mathrm{c}^- $ jumps from negative to positive values, and indicate hence the ``first act'' of the aforementioned $ 0 $--$ \pi $-like transition.

    The ``second act'', such that the junction has undergone the full $ 0 $--$ \pi $-like transition, requires that accordingly $ \varphi_\mathrm{c}^+ $ jumps from positive to negative values, which is indicated by the cusps that occur in the $ I_\mathrm{c}^+ $-curves. 
    Interestingly, while both ``acts'' of the $ 0 $--$ \pi $-like transition happen for weak SOC (nearly) at the same magnetic exchange and might be well observable in experiments, strong $ \lambda_\mathrm{SOC} $ requires a much larger $ \lambda_\mathrm{XC} $ to drive the $ 0 $--$ \pi $-like transition. This could make it experimentally more challenging to study $ 0 $--$ \pi $-like transitions in 2DEGs with extraordinarily strong SOC or small Fermi energy. 
    The physical reason is the rather intricate interplay between the ``quasi-sinusoidal'' and $ \varphi_0 $-shifted CPRs of transverse channels with~$ |k_y| $ smaller and larger than $ |k_y^\mathrm{crit.}| $, respectively; see discussion in~Sec.~\ref{Sec_BoundStates_Magnetic} and~Fig.~\ref{Fig_BoundStatesCurrent_MAGNETIC}.

    The figure of merit distinguishing $0$-like junction states from their 
    $\pi$-like counterparts~[see, e.g., Fig.~\ref{Fig_BoundStatesCurrent_COLORMAP}(c)] is the amount of the area enclosed by the total $I_\mathrm{J}(\varphi)$-curve and the positive $\varphi$-axis, i.e., $A=\int_0^\pi \mathrm{d}\varphi \, I_\mathrm{J}(\varphi)$. 
    Let us focus on the separate contributions to that area coming from the channels with $ |k_y| $ smaller and larger than $|k_y^\mathrm{crit.}| $, respectively, for junctions in the $0$-like state~($\lambda_\mathrm{SOC}=0.4$, $\lambda_\mathrm{XC}=1.5$), Fig.~\ref{Fig_BoundStatesCurrent_MAGNETIC}, and $\pi$-like state
    ($\lambda_\mathrm{SOC}=0.4$, $\lambda_\mathrm{XC}=2.5$), Fig.~\ref{Fig_BoundStatesCurrent_STRONGLY_MAGNETIC}. 
    Looking at the channels' resolved ``quasi-sinusoidal'' CPRs for $|k_y|<|k_y^\mathrm{crit.}| \approx 0.68 k_\mathrm{F}$, we observe that their contributions to $A$ are negative in both studied cases~(and become even more negative when $\lambda_\mathrm{XC}$ is enhanced from $1.5$ to $2.5$), i.e.,~their CPRs always favor negative critical Josephson currents at positive phase differences. 
    Contrarily, the behavior of the $\varphi_0$-shifted CPRs of the  $|k_y|>|k_y^\mathrm{crit.}| \approx 0.68 k_\mathrm{F}$-channels is
    different at $\lambda_\mathrm{XC}=1.5$ and $\lambda_\mathrm{XC}=2.5$. In the first case, Fig.~\ref{Fig_BoundStatesCurrent_MAGNETIC}, they contribute to 
    $A$ positively, i.e.,~they counteract the ``quasi-sinusoidal'' modes 
    and hence favor positive critical Josephson currents at positive phase differences. In total, these positive current contributions dominate over the ``quasi-sinusoidal'' modes, and the junction is overall still in the $0$-like state.
    For $\lambda_\mathrm{XC}=2.5$, Fig.~\ref{Fig_BoundStatesCurrent_STRONGLY_MAGNETIC}, the $\varphi_0$-shifted modes lower their positive contributions to $A$~[effectively their anomalous $\varphi_0$-phases are shifted and the nodes of their individual $I_\mathrm{J}(\varphi)$-curves move closer to $\varphi=\pi/2$] and the junction switches to the $\pi$-like state. 
    Knobs to control the $ 0 $--$ \pi $-like transitions are the SOC strength, which sets the magnitude of $ |k_y^\mathrm{crit.}| $ according to~Eq.~\eqref{Eq_Critky} and hence the weights of ``quasi-sinusoidal'' and $\varphi_0$-shifted modes, and the magnetic exchange strength that controls the $ \varphi_0 $-shifts of the $ |k_y| > |k_y^\mathrm{crit.}| $-channels.

    \begin{figure}
        \centering
        \includegraphics[width=0.49\textwidth]{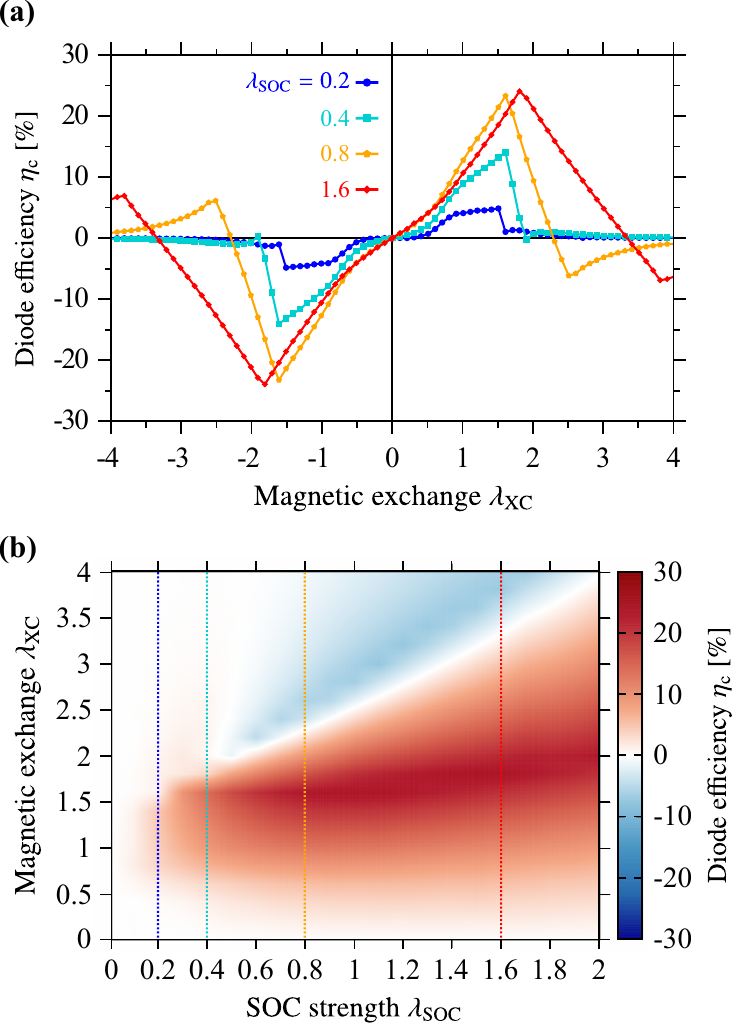}
        \caption{(a)~Calculated Josephson SDE efficiencies~$\eta_\mathrm{c}$~[defined by~Eq.~(\ref{Eq_DiodeEff})] as functions of the magnetic exchange~$ \lambda_\mathrm{XC} $ and for indicated values of Rashba~$ \lambda_\mathrm{SOC} $; the other junction parameters are the same as in~Fig.~\ref{Fig_CritCurrent_DifferentSOC_ReducedSize}. 
        (b)~Same as in~(a), but as color map with a denser sampling of $ \lambda_\mathrm{SOC} $ and $ \lambda_\mathrm{XC} $. 
        The cuts, indicated by the dashed colored lines, correspond to the curves displayed in~(a). }
        \label{Fig_Diode_DifferentSOC_ReducedSize}
    \end{figure}

    To quantify the efficiency of the SDE of the 2DEG-based magnetic Josephson junctions, we evaluate 
    \begin{equation}
        \eta_\mathrm{c}=\frac{\Delta I_\mathrm{c}}{I_\mathrm{c}(0)} = \frac{I_\mathrm{c}^+ - |I_\mathrm{c}^-|}{I_\mathrm{c}(0)} ,
        \label{Eq_DiodeEff}
    \end{equation}
    which measures the difference between the magnitudes of positive and negative critical currents at generic exchange~$ \lambda_\mathrm{XC} $ relative to the critical current~$I_\mathrm{c}(0)$
    of the nonmagnetic junction. 
    Figure~\ref{Fig_Diode_DifferentSOC_ReducedSize}(a) shows $ \eta_\mathrm{c} $ as a function of~$ \lambda_\mathrm{XC} $ and for various SOC strengths~$ \lambda_\mathrm{SOC} $. 
    It is remarkable that---despite the minimality of our model---the calculated diode efficiencies convincingly reproduce all experimentally determined characteristics reported for Al-gated InAs 2DEG-based Josephson junctions in Refs.~\cite{Baumgartner2022a,Baumgartner2022,Costa2023,*Costa2023NatNano}: (i)~a weak ``kink'' (soft steepness) at small~$ \lambda_\mathrm{XC} $, (ii)~followed by a nearly linear increase that reaches (iii)~a maximum~(in terms of a sharp peak at realistic SOC), and (iv)~a down-turn~(that can even be accompanied by a sign reversal at large-enough $\lambda_\mathrm{SOC}$), with a final fading at large $ \lambda_\mathrm{XC} $.

    Based on our critical-current analyses, we can describe these features in terms of their physical origin. 
    The sharp peaks---feature~(iii)---stem from the cusps experienced by~$ |I_\mathrm{c}^-| $, which play the crucial role in the ``first act'' of the $ 0 $--$ \pi $-like transition; since $ |I_\mathrm{c}^-| $ drops into a local minimum, $\eta_\mathrm{c}$~[see Eq.~\eqref{Eq_DiodeEff}] maximizes. 
    Analogously, the diode efficiency gets down-turned---feature~(iv)---around the $ I_\mathrm{c}^+ $-cusps, which indicate the ``second act'' of the $ 0 $--$ \pi $-like transition. 
    At large-enough SOC, $ \eta_\mathrm{c} $ even reverses its sign when further increasing $ \lambda_\mathrm{XC} $. As explained in detail in a recent work~\cite{Lotfizadeh2023}, these sign changes are not necessarily a signature of the $ 0 $--$ \pi $-like~transition, which agrees well with our numerical results. For example, at $ \lambda_\mathrm{SOC} = 0.4 $, $ \eta_\mathrm{c} $ does not experience a (visible) sign change, whereas the CPRs shown in~Fig.~\ref{Fig_BoundStatesCurrent_COLORMAP}(c) undoubtedly reflect a $ 0 $--$ \pi $-like transition. 
    Instead, the $ \eta_\mathrm{c} $-sign reversals were attributed to an asymmetry in the $ I_\mathrm{c}^\pm $--$ \lambda_\mathrm{XC} $~relations, i.e., $ |I_\mathrm{c}^\pm(\lambda_\mathrm{XC})| \neq |I_\mathrm{c}^\pm(-\lambda_\mathrm{XC})| $~[also visible in~Figs.~\ref{Fig_CritCurrent_DifferentSOC_ReducedSize}(a) and \ref{Fig_CritCurrent_DifferentSOC_ReducedSize}(b)], in~Ref.~\cite{Lotfizadeh2023}, serving as another fingerprint of the simultaneous breakdown of space-inversion and time-reversal symmetries. Inspecting Figs.~\ref{Fig_CritCurrent_DifferentSOC_ReducedSize}(a) and \ref{Fig_CritCurrent_DifferentSOC_ReducedSize}(b), the maximal critical-current amplitudes as functions of~$ \lambda_\mathrm{XC} $ occur at finite~$ |\lambda_\mathrm{XC}^*| $. Contrarily, if SOC were absent, the critical currents~$ |I_\mathrm{c}^\pm| $ would always be symmetric with respect to the sign of~$ \lambda_\mathrm{XC} $ with their maxima at~$ \lambda_\mathrm{XC} = 0 $. 
    As the shift~$ |\lambda_\mathrm{XC}^*| $ increases with the SOC~strength~$ \alpha $, the $ I_\mathrm{c}^\pm $--$ \lambda_\mathrm{XC} $~relations become more asymmetric with stronger SOC. The authors of Ref.~\cite{Lotfizadeh2023} predicted a nearly linear dependence of $ |\lambda_\mathrm{XC}^*| $ on~$ \alpha $, which can be used to determine the SOC strength from experimental measurements of the critical currents. 
    From Figs.~\ref{Fig_CritCurrent_DifferentSOC_ReducedSize}(a) and \ref{Fig_CritCurrent_DifferentSOC_ReducedSize}(b), we see that the $ I_\mathrm{c}^+ $- and $ |I_\mathrm{c}^-| $-curves may eventually cross due to these asymmetric $ I_\mathrm{c}^\pm $--$ \lambda_\mathrm{XC} $~relations, what indicates the sign reversal of~$ \eta_\mathrm{c} $.

    The weak ``kink'' and the afterwards nearly linear increase of the diode efficiency at smaller $\lambda_\mathrm{XC}$~[features (i) and (ii)] are again related to the qualitatively different CPRs of the individual $ k_y $-channels. 
    The slope of the ``kink'' rises with increasing $\lambda_\mathrm{SOC}$; see
    Fig.~\ref{Fig_Diode_DifferentSOC_ReducedSize}(a). As explained before,
    larger $\lambda_\mathrm{SOC}$ means lower $|k_y^\mathrm{crit.}|$ and the contributions of the $\varphi_0$-shifted modes become more dominant. At small magnetic exchange, these $ \varphi_0 $-shifts imprint a nearly linear $ \lambda_\mathrm{XC} $-dependence on~$ \eta_\mathrm{c} $~\cite{Buzdin2008}.

    To cover the whole range of junction parameters, Fig.~\ref{Fig_Diode_DifferentSOC_ReducedSize}(b) shows the diode efficiency as a function of both the SOC and magnetic exchange strengths. 
    The results confirm our previous observations, covering the maximal $ \eta_\mathrm{c} $-peaks of about~$ 30 \, \% $ diode efficiency and the (possible) sign change of $ \eta_\mathrm{c} $ at elevated SOC. 
    Apart from the good qualitative agreement, also the order of magnitude of the Josephson SDE extracted from our theoretical model compares reasonably well with the experimental values~(about~$ 20 \, \% $) measured in Al-gated InAs 2DEG-based junctions~\cite{Baumgartner2022a,Baumgartner2022,Costa2023,*Costa2023NatNano}.

    \begin{figure}
        \centering
        \includegraphics[width=0.49\textwidth]{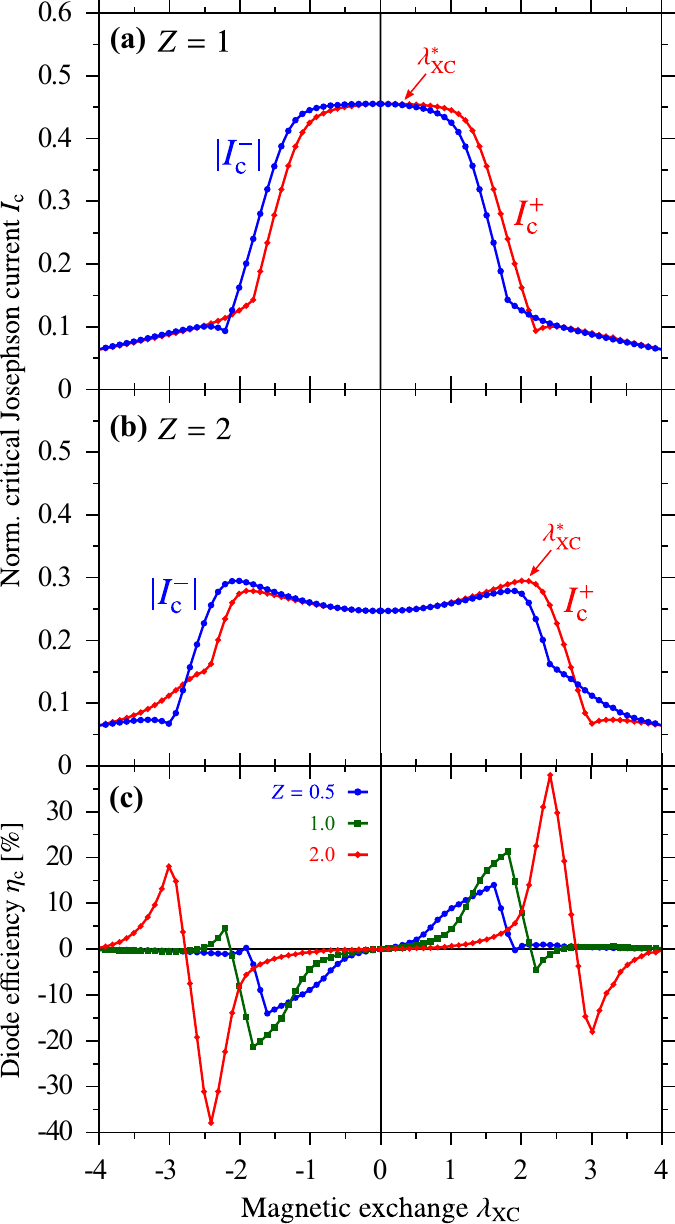}
        \caption{Polarity dependence of the critical Josephson currents~$ I_\mathrm{c}^+ $ and $ |I_\mathrm{c}^-| $, displayed as functions of the magnetic exchange~$ \lambda_\mathrm{XC} $ and for~$ \lambda_\mathrm{SOC} = 0.4 $; the barrier $ Z $-parameters are (a)~$ Z = 1 $ and (b)~$ Z = 2 $. The shift~$ \lambda_\mathrm{XC}^* $ of the critical-current maxima strongly depends on the junction transparency through~$ Z $. 
        (c)~Calculated Josephson SDE efficiencies~$\eta_\mathrm{c}$~[defined by~Eq.~(\ref{Eq_DiodeEff})] as functions of the magnetic exchange~$ \lambda_\mathrm{XC} $ for $ \lambda_\mathrm{SOC} = 0.4 $ and indicated $ Z $-values of the tunnel barrier. }
        \label{Fig_CritCurrent_DifferentBarriers_ReducedSize}
    \end{figure}

    Another knob to tune the Josephson SDE---except for the SOC---is the $ Z $-parameter of the delta-like tunnel barrier, which is experimentally determined by the transparency of the nonsuperconducting weak link. 
    In all calculations discussed so far, we assumed a rather small value of~$ Z = 0.5 $ to model the highly transparent junctions~(transparency~$ \overline{\tau} = 1 / [1 + (Z/2)^2] \approx 0.94 $) corresponding to recent experiments~\cite{Baumgartner2022a,Baumgartner2022,Costa2023,*Costa2023NatNano}. 
    However, Fig.~\ref{Fig_CritCurrent_DifferentBarriers_ReducedSize}(c) suggests that it might be worth to focus on less-transparent junctions in future experiments, as some of the most striking SDE features---i.e., the sharp $ \eta_\mathrm{c} $-peaks and, more interestingly, the SDE sign reversals---appear then in a more prominent way. 
    As explained above, the sign reversals of~$ \eta_\mathrm{c} $ are related to the asymmetric $ |\lambda_\mathrm{XC}^*| $-shift of the critical-current maxima. Our calculations shown in~Figs.~\ref{Fig_CritCurrent_DifferentBarriers_ReducedSize}(a) and \ref{Fig_CritCurrent_DifferentBarriers_ReducedSize}(b) suggest that $ |\lambda_\mathrm{XC}^*| $ does not only depend on the SOC strength~$ \alpha $ as explored in~Ref.~\cite{Lotfizadeh2023}, but even more strongly on the barrier parameter~$ Z $. 
    Larger~$ Z $~(reduced interfacial transparency) results in a greatly enhanced critical-current asymmetry and more clearly apparent sign changes of~$ \eta_\mathrm{c} $. 
    From that viewpoint, one reason that these sign changes were hardly discernible in the experiment~\cite{Baumgartner2022a,Baumgartner2022,Costa2023,*Costa2023NatNano} might be that the junction transparency was even larger than the theoretically assumed~$ \overline{\tau} \approx 0.94 $.

    \section{Conclusions        \label{Sec_Conclusions}}

    To summarize, we formulated a minimal theoretical model capturing the spectral---Andreev bound states---and transport---Josephson current---characteristics of 2DEG-based ballistic Josephson junctions with ultrathin ferromagnetic weak links. 
    Investigating a wide range of realistic junction parameters, we analyzed the impact of magnetic exchange, SOC, and junction transparency on the underlying Josephson CPRs---scrutinizing their distortion, anomalous $ \varphi_0 $-phase shift, Josephson SDE, and $ 0 $--$ \pi $-like transitions. 
    We quantified the diode effect in terms of the SDE efficiency~$ \eta_\mathrm{c} $ that can reach values beyond~$ 30 \, \% $, and demonstrated that $ \eta_\mathrm{c} $ depends on the magnetic exchange~$ \lambda_\mathrm{XC} $ in a unique way, which agrees reasonably well with recent experiments. 
    Our model allows to unveil current-reversing $ 0 $--$ \pi $-like transitions or extract the magnitude of the SOC from experimental measurements of~$ \eta_\mathrm{c} $. 
    Finally, we showed that the SDE can be even further enhanced through reducing the transparency of the weak link. This might open a promising path for future experimental works provided that mostly junctions with high transparency have been investigated so far.

    \begin{acknowledgments}
        A.C. and J.F. acknowledge funding by the Deutsche Forschungsgemeinschaft~(German Research Foundation) within the Research grant ``Spin and magnetic properties of superconducting tunnel junctions''---Project-ID 454646522. 
        D.K.~acknowledges partial financial support from the project IM-2021-26 (SUPERSPIN) funded by Slovak Academy of Sciences via the programme IMPULZ 2021. 
    \end{acknowledgments}

    \bibliography{paper}

\end{document}